\documentclass[aps,prl,twocolumn,floatfix,groupedaddress,superscriptaddress,footinbib,notitlepage,showpacs]{revtex4-2}
\usepackage{graphicx,graphics,times,bm,bbm,bbold,amssymb,amsmath,amsfonts,dsfont,hyperref,color,geometry, longtable}
\usepackage{algorithm}
\usepackage{algorithmic}
\usepackage[english]{babel}
\usepackage[dvipsnames]{xcolor}
\usepackage{url}
\usepackage{physics}
\usepackage{cleveref}
\usepackage[caption=false]{subfig}  
\usepackage{float}
\usepackage{tikz}
\usetikzlibrary{fit,arrows.meta}
\hypersetup{colorlinks,linkcolor=blue,citecolor=blue,urlcolor=blue}

\newtheorem{lemma}{Lemma}
\newtheorem{theorem}{Theorem}
\newtheorem{definition}{Definition}

\usepackage{natbib}
\usepackage[autostyle]{csquotes}
\usepackage{relsize}
\usepackage{soul}
\usepackage{comment}
\newenvironment{proof}[1][Proof]{\begin{trivlist}
		\item[\hskip \labelsep {\bfseries #1}]}{\end{trivlist}}
\newcommand{\qed}{\nobreak \ifvmode \relax \else
	\ifdim\lastskip<1.5em \hskip-\lastskip
	\hskip1.5em plus0em minus0.5em \fi \nobreak
	\vrule height0.5em width0.60em depth0.2em\fi}

\newcommand{\ignore}[1]{}

\newcommand{\re}{\mathrm{e}}

\newcommand{\average}[1]{\left\langle#1\right\rangle}

\newcommand{\qfi}[1]{\mathcal{F}#1}

\newcommand*{\rom}[1]{\expandafter\@slowromancap\romannumeral#1@}

\def\>{\rangle}
\def\<{\langle}
\let\oldsqrt\sqrt
\def\sqrt{\mathpalette\DHLhksqrt}
\def\DHLhksqrt#1#2{%
\setbox0=\hbox{$#1\oldsqrt{#2\,}$}\dimen0=\ht0
\advance\dimen0-0.2\ht0
\setbox2=\hbox{\vrule height\ht0 depth -\dimen0}%
{\box0\lower0.4pt\box2}}
\DeclareFontFamily{OT1}{pzc}{}
\DeclareFontShape{OT1}{pzc}{m}{it}%
              {<-> s * [1.25] pzcmi7t}{}
\DeclareMathAlphabet{\mathpzc}{OT1}{pzc}%
                                 {m}{it}

\hypersetup{colorlinks,linkcolor={blue},citecolor={blue},urlcolor={blue}}
\begin{document}
\title{Many-body $k$-local ground states as probes for unitary quantum metrology}

\author{Majid Hassani}
\thanks{These authors contributed equally to this work.}
\affiliation{Instituut-Lorentz, Universiteit Leiden, P.O. Box 9506, 2300 RA Leiden, The Netherlands}
\affiliation{$\langle aQa^{L}\rangle$ Applied Quantum Algorithms Leiden, The Netherlands}

\author{Mengyao Hu}
\thanks{These authors contributed equally to this work.}
\affiliation{Instituut-Lorentz, Universiteit Leiden, P.O. Box 9506, 2300 RA Leiden, The Netherlands}
\affiliation{$\langle aQa^{L}\rangle$ Applied Quantum Algorithms Leiden, The Netherlands}
\affiliation{Centre for Quantum Technologies, National University of Singapore, 3 Science Drive 2, Singapore 117543, Singapore}

\author{Guillem M\"uller-Rigat}
\thanks{These authors contributed equally to this work.}
\affiliation{Faculty of Physics, Astronomy and Applied Computer Science, Jagiellonian University, ul. \L ojasiewicza 11, 30-348 Kraków, Poland}

\author{Matteo Fadel}
\email{fadelm@phys.ethz.ch}
\affiliation{Department of Physics, ETH Zurich, 8093 Zurich, Switzerland}

\author{Jordi Tura}
\email{tura@lorentz.leidenuniv.nl}
\affiliation{Instituut-Lorentz, Universiteit Leiden, P.O. Box 9506, 2300 RA Leiden, The Netherlands}
\affiliation{$\langle aQa^{L}\rangle$ Applied Quantum Algorithms Leiden, The Netherlands}

\begin{abstract}
Multipartite quantum states saturating the Heisenberg limit of sensitivity typically require full-body interactions to be prepared. On the other hand, experimentally practical Hamiltonians often involve few-body correlators only. Here, we study the metrological performances under this constraint, using tools derived from the quantum Fisher information. 
Our work applies to any encoding generator, also including a dependence on the parameter. 
For moderate system sizes up to hundred spins (even), we find that symmetric ground states of Hamiltonians with significantly few-body interactions exhibit an average quantum Fisher information that agrees with that of Haar-random symmetric states and displays Heisenberg-like scaling sensitivity. We also discuss the interaction requirements to generate metrologically useful Haar-random symmetric eigenstates. Finally, we establish a tradeoff between the Hamiltonian's gap, which quantifies preparation hardness, and the quantum Fisher information of the corresponding ground state. 
\end{abstract}

\date{\today}
\maketitle
\twocolumngrid

The use of a quantum system for estimating an unknown parameter lies at the heart of quantum metrology and sensing \cite{giovannetti_quantum_2006-1,paris_quantum_2009,giovannetti_advances_2011,escher_general_2011,degen_quantum_2017,pezze_quantum_2018,polino_photonic_2020,fadel_quantum_2025}, which form one of the foundational pillars of emerging quantum technologies \cite{simon_introduction_2025}. 
Leveraging quantum resources, such as quantum correlations, can enhance the precision of parameter estimation beyond what is achievable by classical means. 
In parallel with extensive experimental progress \cite{pedrozo-penafiel_entanglement_2020,troullinou_quantum-enhanced_2023,cassens_entanglement-enhanced_2025,pezze_advances_2025}, theoretical investigations, including both analytical \cite{demkowicz-dobrzanski_elusive_2012,demkowicz-dobrzanski_using_2014,hassani_digital_2017,frowis_does_2019,pezze_heisenberg-limited_2020,annabestani_multiparameter_2022,lu_number_2024,kurdzialek_using_2023,hassani_privacy_2025,abiuso_fundamental_2025} and numerical approaches \cite{jarzyna_matrix_2013,chabuda_tensor-network_2020,chabuda_tnqmetro_2022,kurdzialek_quantum_2025,dulian_qmetro_2025}, aim to identify optimal schemes that attain the fundamental limits of precision imposed by quantum mechanics. 

A general quantum metrology protocol can be decomposed into three essential stages: (i) preparation of the probe state on which the parameter of interest is encoded, (ii) actual encoding of the unknown parameter through interaction with the probe, and (iii) measurement of the probe state after encoding to extract information about the parameter \cite{giovannetti_quantum_2006-1}. Here, we focus on the first stage. Determining the best probe is a crucial step, as it sets an upper bound on the precision achievable in the metrological task, which is given by the quantum Fisher information (QFI)~\cite{braunstein_statistical_1994}. 
Equally important, however, is to identify probe states that are feasible to prepare in the laboratory. 
In this direction, we explore the metrological usefulness of ground states of permutation-invariant (PI) Hamiltonians which involve \textit{only} few-body interactions. This class of Hamiltonians include Lipkin-Meshkov-Glick (LMG) models~\cite{lipkin_validity_1965}, which are naturally realized in various platforms, including ultracold atomic ensembles~\cite{kawaguchi_spinor_2012}. Excellent experimental developments also enable the control of the interaction length in many-body systems such as ion crystals \cite{monroe_programmable_2021,britton_engineered_2012,islam_emergence_2013,jurcevic_quasiparticle_2014,katz_programmable_2023}, Rydberg arrays \cite{gambetta_long-range_2020}, and atomic ensembles in optical cavities \cite{periwal_programmable_2021}.
There, ground states can be prepared through adiabatic protocols \cite{evrard_observation_2021,zou_beating_2018} as long as the Hamiltonian gap is not too small \cite{rai_hierarchy_2024,rai_spectral_2025}. In previous works~\cite{oszmaniec_random_2016,imai_metrological_2025}, a connection between Haar random symmetric states and metrological usefulness was highlighted. However, such a connection applies only to the parameter-independent encoding, and it requires Haar random states, which are notoriously difficult to prepare, even in the symmetric space, as they demand maximal complexity, i.e., up to full system size interactions ~\cite{gross_evenly_2007,zhu_clifford_2016}.

In this work, we investigate \textit{typical} metrological properties of ground states of random, physically-motivated Hamiltonians which involve only $k$-local interactions, with $k$ even. As a result, we show that, on average, these physically accessible ground states deliver close to maximal sensitivity, displaying Heisenberg-like scaling $\mathrm{QFI}\sim O(N^2)$, recovering the performance of Haar-random symmetric states~\cite{oszmaniec_random_2016,imai_metrological_2025}~for systems of about hundred spins. We further reveal a tradeoff relation between the Hamiltonian’s gap and the metrological usefulness of the corresponding ground state.
We demonstrate that our results also apply to the case in which the encoding generator explicitly depends on the parameter, which remains much less explored, both in theory~\cite{pang_quantum_2014,pang_optimal_2017,annabestani_multiparameter_2022} and experiment~\cite{souquet-basiege_quantum_2025,alderete_nonlinear_2025}.\\

\textbf{Preliminaries}---In order to estimate the unknown parameter $\theta$, one can prepare a probe quantum state, $\rho_{0}$, on which the parameter is imprinted as $\rho_{0} \rightarrow \rho_{\theta}$ via quantum dynamics. 
Performing a measurement on the evolved probe state, $\rho _{\theta}$, provides information about $\theta$. 
In quantum metrology, the fundamental bound on the attainable precision in estimating the unknown parameter $\theta$ for a given statistical model $\rho _{\theta}$, can be addressed by the Cramér-Rao bound \cite{braunstein_statistical_1994, braunstein_generalized_1996,giovannetti_quantum_2006-1,paris_quantum_2009}
\begin{equation}\label{Cramer-Rao}
\delta\theta\geqslant\frac{1}{\sqrt{\qfi[\theta]}},
\end{equation}
in which $\delta\theta$ represents the estimation error and $\qfi[\theta]$ is the QFI which quantifies the maximum amount of extractable information about the unknown parameter $\theta$ over \textit{all} possible measurements and is a function of the final state $\rho_\theta$. 
In this work, we focus on pure probes $\rho_\theta = \ketbra{\psi_\theta}$ that depend continuously on $\theta$. Under this condition, the QFI can be expressed as
\begin{equation}
\label{QFI-pure-states}
  \qfi[\ket{\psi_\theta}]= 4(\average{\partial_{\theta}\psi_{\theta}|\partial_{\theta}\psi_{\theta}}-\average{\psi_{\theta}|\partial_{\theta}\psi_{\theta}}\average{\partial_{\theta}\psi_{\theta}|\psi_{\theta}}).
\end{equation}

Importantly, if the parameter encoding results from a unitary evolution, i.e. $\ket{\psi_\theta}= \mathbf{U}_\theta \ket{\psi_0}$, we can write $\ket{\partial_\theta\psi_\theta}= -i \mathbf{K}_\theta \ket{\psi_\theta}$, with the Hermitian operator $\mathbf{K}_\theta= i(\partial_{\theta}\mathbf{U}_{\theta})~\mathbf{U}_{\theta}^{\dagger}$.
Equation~\eqref{QFI-pure-states} thus simplifies to
\begin{equation} \label{QFI-pure-states-unitary}
    \qfi[\ket{\psi_\theta}]= 4\Delta ^{2} \mathbf{K}_{\theta} = 4(\bra{\psi_\theta}\mathbf{K}_{\theta}^{2}\ket{\psi_\theta}-\bra{\psi_\theta}\mathbf{K}_{\theta}\ket{\psi_\theta}^{2}).
\end{equation}
Furthermore, we consider a system composed of $N$-qubits in which the parameter is imprinted identically in each qubit, $\mathbf{U}_\theta = U_{\theta}^{\otimes N}$, with $U_\theta = e^{-i G_\theta}$ in which $G_{\theta}$ denotes the generator of the unitary evolution. For such a case,
\begin{equation}\label{QFI_PDH_Variance_upper_bound}
    \qfi[\ket{\psi_\theta}]\leqslant N^{2}\Vert K_{\theta}\Vert_{\text{sn}}^{2},
\end{equation}
where $\Vert K_{\theta}\Vert _{\text{sn}}=\lambda _{M}(K_{\theta})-\lambda _{m}(K_{\theta})$ denotes the operator seminorm of Hermitian operator $K_\theta= i(\partial_{\theta}U_{\theta})~U_{\theta}^{\dagger}$, and $\lambda_M/m$ represents the max/min eigenvalue, see the Supplemental Material (SM) for derivation. Equation.~\eqref{QFI_PDH_Variance_upper_bound} can be saturated by the GHZ-type state $\ket{\psi_\theta} = (\ket{\lambda_M(\theta)}^{\otimes N} +\ket{\lambda_m(\theta)}^{\otimes N})/\sqrt{2}$, where $\ket{\lambda_{M/m}(\theta)} $ are respectively the max/min eigenvectors of $K_{\theta}$, which is so-called the Heisenberg limit (HL), $\qfi[\ket{\psi_\theta}]=N^2$. In addition, whenever the QFI surpasses the standard quantum limit (SQL), $\qfi[\ket{\psi_\theta}]>N$, one can say that the quantum state $\rho_\theta$ allows for quantum-enhanced sensitivity through the presence of metrologically useful entanglement, since a QFI saturating the SQL can be achieved by classical means. Unlike the ground states considered here, GHZ states are particularly demanding to prepare and because they require full-body interactions.

In order to analyze the behavior of the QFI with the number of particles $N$, we face an exponential growth of the Hilbert space dimension. To address this challenge, we derive an expression for the QFI for pure probe states $\ket{\psi_0}^{\rm sym.}\in \mathcal{S}[(\mathbb{C}^2)^{\otimes N}]$ belonging to the symmetric subspace $\mathcal{S}[(\mathbb{C}^2)^{\otimes N}]$, which is spanned by the Dicke states $\{\ket{D_N^{n}} \propto \ket{0}^{\otimes n}\otimes \ket{1}^{\otimes (N-n)} + \mathrm{perms.} \}$, where $\mathrm{perms.}$ indicates unique permutations~\cite{marconi_symmetric_2025}. Under these restrictions, the QFI explicitly reads: 
   \begin{align}\label{QFI-symmetric-subspace_maintext}
    \qfi[\ket{\psi_{\theta}}^{\text{sym.}}]&=4 \sum_{m,n,l=0}^{N}\alpha _{m}^{*}\alpha _{n}(\partial _{\theta}\mathpzc{C}_{lm})^{*}~\partial _{\theta}\mathpzc{C}_{ln}\nonumber\\
    &~~~-4\Big\vert \sum_{m,n,l=0}^{N}\alpha _{m}^{*}\alpha _{n}\mathpzc{C}_{lm}^{*}~\partial _{\theta}\mathpzc{C}_{ln}\Big\vert ^{2},
\end{align} 
where $\alpha_n = \bra{D^n_N}\ket{\psi_0}^{\rm sym.}$ are the components of $\ket{\psi_0}^{\rm sym.}$ in the Dicke basis and $\mathpzc{C}_{nm} = \bra{D^n_N}\mathbf{U}_\theta\ket{{D^m_N}}$, see the SM. Equation~\eqref{QFI-symmetric-subspace_maintext} implies that the QFI can be directly computed in the symmetric subspace. 
By working within the symmetric subspace, which has dimension $(N+1)$, we drastically reduce the computational complexity, enabling efficient analysis even for large $N$. Even though the symmetric subspace is exponentially small compared to the total Hilbert space, $(\mathbb{C}^2)^{\otimes N}$, those states maximally metrologically useful are, in fact, symmetric~\cite{pezze_quantum_2018, oszmaniec_random_2016}. In the rest of our work, we focus on this scenario and omit the label ``sym.'' for simplicity of notation.

\vspace{2mm}
\textbf{System}---As already mentioned, we aim to characterize the metrological usefulness of ground states of experimentally practical Hamiltonians. In this way, we define the system's Hamiltonian as follows
\begin{equation}\label{H_sys}
    H= \mathlarger{\mathlarger{\sum}}_{ a+b+c\leqslant k}\Gamma_{abc}~S_{abc},
\end{equation}

where $\Gamma_{abc}\in \mathbb{R}$ denotes the coefficients, and the operator $S_{abc}$ represents all-to-all PI $k$-body interactions and is given by
\begin{equation}
    \label{def_k_body_corr}
S_{abc} = \frac{1}{2^{k}k!}(\sigma_x^{\otimes a}\otimes \sigma_y^{\otimes b}\otimes \sigma_z^{\otimes c}\otimes \mathbb{1}^{\otimes(N-k)} + \mathrm{perms.}),
\end{equation}
in which $\sigma_{x,y,z}$ are the standard Pauli matrices, $\mathbb{1}$ is the identity matrix, and $\mathrm{perms.}$ indicates all particle permutations. Physical Hamiltonians usually involve only few-body interactions (low $k$).  In this work, we consider ground states in the symmetric subspace. Such states are not necessarily the ground states of the full Hamiltonian, as the true ground state may belong to a permutation-invariant (PI) sector other than the fully symmetric one. Next, we derive a recursive formula for computing $k$-body interactions from lower-order ones
    \begin{align}\label{recursive_k-body_correlators-PI}
        S_{abc}&=\frac{1}{k}S_{ab(c-1)}S_{001}- \frac{(N-k+2)(c-1)}{4k(k-1)} S_{ab(c-2)}\nonumber\\
        &~~~+\frac{i}{2k}(aS_{(a-1)(b+1)(c-1)}-b S_{(a+1)(b-1)(c-1)}).
    \end{align}
From Eq.~(\ref{recursive_k-body_correlators-PI}), $H$ can be described efficiently in the symmetric subspace using dynamic programming techniques and its ground state can be computed for large system sizes $N$, see the SM. The associated QFI can then be evaluated efficiently using Eq.~\eqref{QFI-symmetric-subspace_maintext}, with computational complexity $O(N^{4}+k^{3}N^{3})$, see also the SM for a detailed discussion. This versatile computational tool has potential applications beyond quantum metrology, including the construction of Bell inequalities based on few-body correlators~\cite{tura_detecting_2014,tura_nonlocality_2015,aloy_deriving_2024, muller-rigat_inferring_2021, muller-rigat_three-outcome_2024}.  \\

\textbf{Results}---We begin with the case in which the parameter is encoded linearly as a phase, and conventionally taken as the spin operator along the $x$-direction in which $G_\theta =\theta \sigma_x /2$. As a first example, let us consider two-body correlators, $k=2$, and a Hamiltonian, Eq.~\eqref{H_sys}, with only nonzero coefficient $ \Gamma_{002} = \gamma$, i.e.   
\begin{equation}
\label{eq:H_squeezing}
    \tilde{H} = -\vec{S}^2 + \gamma S_{002} .
\end{equation}
This Hamiltonian is an instance of the LMG model~\cite{lipkin_validity_1965}. For $N$ even and $\gamma >0$, the ground state of Eq.~\eqref{eq:H_squeezing} is the balanced Dicke state,  $\ket{D^{N/2}_{N}}$, which display exceptional metrological properties with QFI equal to $N(N/2 +1)$. For $N$ odd and/or $\gamma \leqslant 0$, the ground state of $\tilde{H}$ is degenerated also in the symmetric subspace. While the ground state subspace may contain metrologically useful states, distinguishing them from the less useful ones generally requires experimentally demanding many-body interactions to lift the degeneracy. As an example, for $\gamma<0$ and $N$ even, the ground subspace contains the classically correlated states $\ket{0}^{\otimes N}, \ket{1}^{\otimes N}$, as well as the GHZ, $(\ket{0}^{\otimes N} + \ket{1}^{\otimes N})/\sqrt{2}$, which is maximally metrologically useful, see discussion below Eq.~\eqref{QFI_PDH_Variance_upper_bound}. However, to single out the GHZ state, one needs to add $N$-body interactions $S_{N00}=\sigma_x^{\otimes N}$ to break the degeneracy. The metrological usefulness of \textit{subspaces} can be certified with the tools developed in Ref.~\cite{muller-rigat_certifying_2023}, see also Ref.~\cite{frerot_symmetry_2024}. Here, we may focus only on the cases where the ground state is unique in the symmetric subspace, as it is guaranteed that it can be prepared from few-body observables alone.

\begin{figure}[h!]
    \centering
     \subfloat[ Average and fluctuations of the QFI ]{\includegraphics[width=\linewidth]{  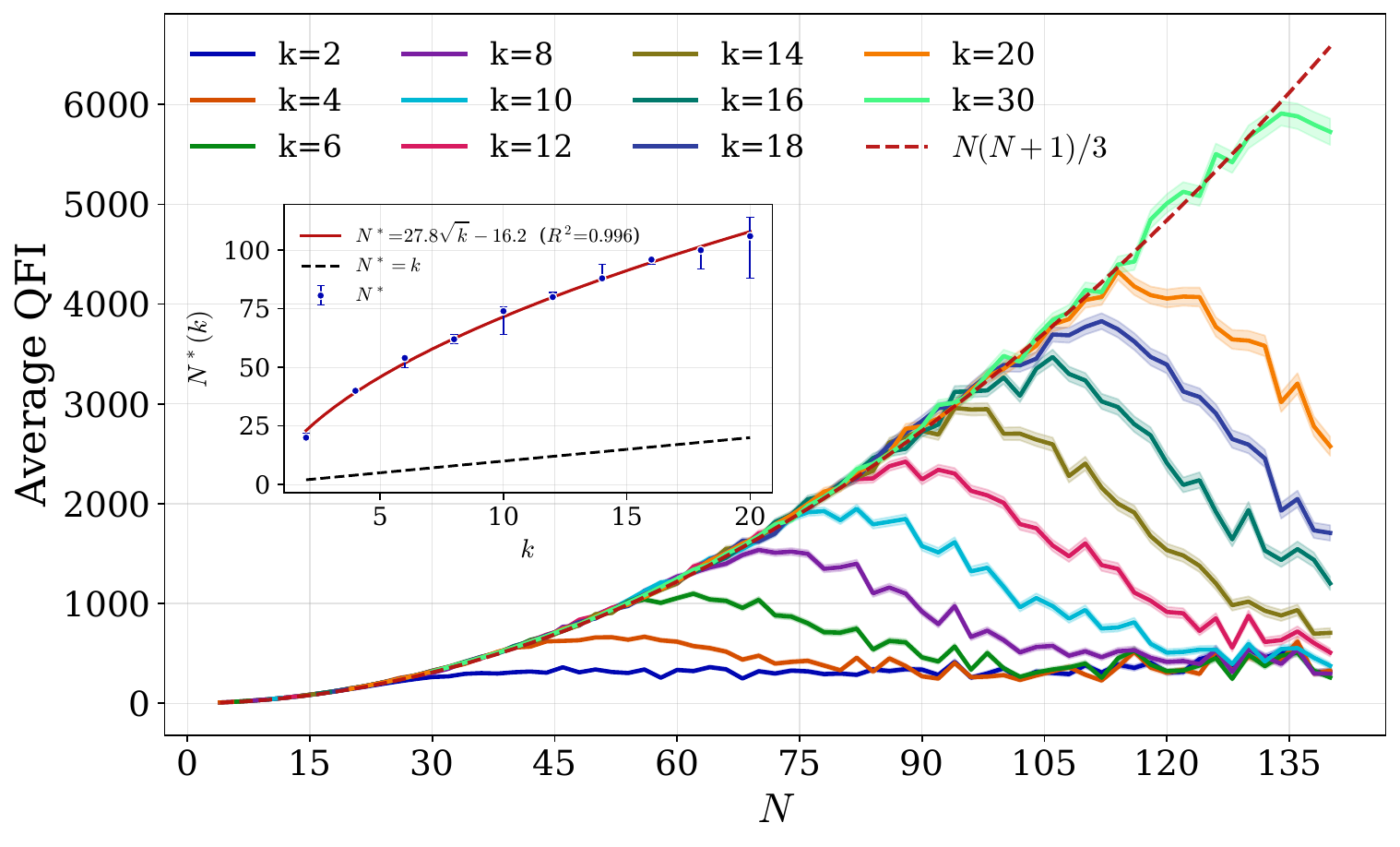}} \\
    \subfloat[The QFI distribution when $k=4$]{\includegraphics[width=\linewidth]{  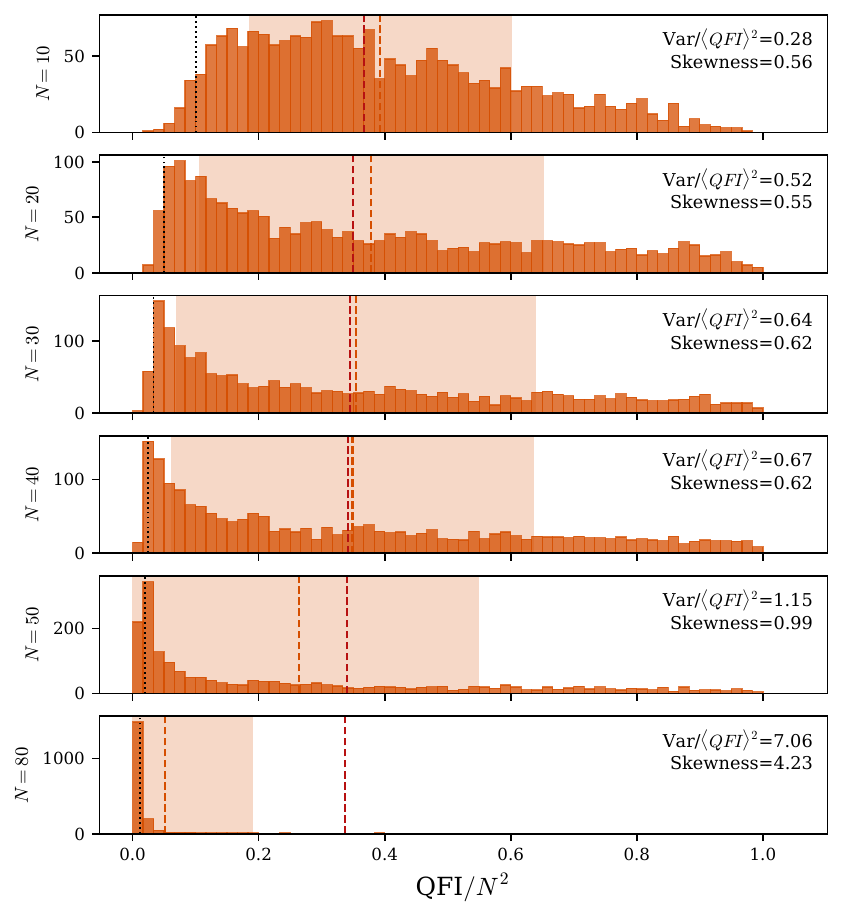}}
    \caption{$(a)$ Average and fluctuations of the QFI as a function of $N$ (even) over 2000 ground states of $H$, constructed with randomly sampled coefficients $\Gamma _{abc}$ drawn from the normal distributions $\mathcal{N}(0,k!/(a!b!c!))$ defined in the text, for various $k$-body interactions. Solid curves denote sample averages; shaded regions indicate $\pm$ standard error of the mean (SEM). The red dashed line shows the reference value $N(N+1)/3$. 
 Inset: critical $N$ below which the average QFI is compatible with Heisenberg-like scaling $N(N+1)/3$. The solid line is the fit $N^{*}=27.8\sqrt{k}-16.2$ with fit quality $R^{2}=0.996$ (in least-squares fitting). The dashed line is $N^{*}=k$.
    (b) Frequency histograms of the QFI for $k=4$ and $N=10,20,30,40,50,80$. 
     In each panel, the orange dashed
line marks the sample average $\langle \mathrm{QFI}\rangle$, the shaded
band spans one standard deviation about the mean,
$\langle \mathrm{QFI}\rangle \pm \sigma$, and the dark-red dashed line
marks the reference value $\mathrm{QFI}=N(N+1)/3$, the dotted line marks the SQL.
}
    \label{H_X-random_sampling}
\end{figure}

In this work, we investigate how generic is metrologically useful entanglement among ground states of random physical Hamiltonians involving exactly $k$-body interactions. To that end, we sample $\Gamma_{abc}$, with $a+b+c = k$, randomly from independent normal distributions $\mathcal{N}(0,\sigma_{abc}^{2})$ of zero mean and variance $\sigma_{abc}^{2}=k!/(a!b!c!)$, the number of distinct orderings of the $k$ Pauli matrices in Eq.~\eqref{def_k_body_corr}. Since the Hilbert--Schmidt norm of $S_{abc}$ is proportional to $\sqrt{a!b!c!/k!}$, this choice gives every term of Eq.~\eqref{H_sys} the same weight. Note that similar ensembles of random Hamiltonians, such as the embedded Gaussian unitary ensemble~\cite{kota_embedded_2014}, have previously been investigated in the context of random matrix theory. Next, we analyze the distribution of the QFI across these random ground states, thereby characterizing the metrological performance expected in disordered systems with uncontrolled interaction strengths. We find that for $N$ odd and $k$ even the ground states of typical Hamiltonians of the form Eq.~\eqref{H_sys} are doubly degenerated. This follows from Kramers' theorem~\cite{kramers_theorie_1930}, which affirms that the spectra of Hamiltonians is doubly degenerated for systems of half-integer total spin and invariant under time reversal $\vec{S}\mapsto -\vec{S}$. For odd $k$, while the ground state is unique, our numerical exploration shows usefulness compatible with the SQL for low $k$. We now turn to the case where both $N$ and $k$ are even, for which the conclusion changes. The results are shown in Fig.~\ref{H_X-random_sampling}.

We observe that the average value of the QFI over even few-body interactions and even number of particles $N$ is compatible with $N(N+1)/3$  (Heisenberg-like scaling) up to a critical system size $N^{*}(k)$ that grows with the interaction order and reaches $N^{*}\approx 100$ spins for $k\approx 20$, Fig.~\ref{H_X-random_sampling}(a).
This scaling was posed in Refs.~\cite{oszmaniec_random_2016,imai_metrological_2025} as the average QFI over symmetric states distributed according to the Haar measure. Conversely, as proven in these works, generic Haar multipartite states do not surpass the SQL, which motivates our choice of Hamiltonians with symmetric ground state of Eq.~\eqref{H_sys}. Symmetric Haar-random states are notably difficult to produce as full-body interactions $k=N$ are needed for $H$ to span the symmetric subspace. Indeed, interactions involving up to $k$-body are polynomials of degree at most power $k$ in the collective spin $S_{001}, S_{010}, S_{100}$. They therefore span the space of multipole operators of rank at most $k$, whose dimension is $(k+1)^2$. In turn, the space of fully symmetric Hamiltonians has dimension $(N+1)^2$, and sampling from this space must be performed in such a way that all ground states are equally probable. The total number of $k$-body interactions up to $N$  appearing in Eq.~(\ref{H_sys}) is $\sum _{k=0}^{N}\binom{k+2}{2}\in {O}(N^{3})$. In the End Matter, we discuss how to choose the $(N+1)^2$ interactions $S_{abc}$, which necessarily include full-body $k=N$ correlators, and how to sample the corresponding couplings $\Gamma_{abc}$ such that the ground states of the resulting Hamiltonians are Haar-random. Note that the sampling procedure used in Fig.~\ref{H_X-random_sampling}, which involves only low-order correlations with $k\ll N$, is different from this procedure. Interestingly, it nevertheless yields results consistent with the Haar-random prediction.

The inset of Fig.~\ref{H_X-random_sampling} (a) shows, for each locality $k$, the critical number of particles, $N^{*}$, below which the average QFI agrees with the Haar-random value. Such Heisenberg-like scaling of the QFI can be achieved for system sizes significantly larger than $k$, see the dashed line corresponding to $N = k$. 
In particular, for $k\leqslant 20$, we find that the growth of the critical $N^*$ is well described by a square-root law. Moreover, we observe that for moderate system sizes ($N^* \approx 100$), $N^*$ increases monotonically with $k$, indicating that the interaction locality $k$ serves as a resource for enhancing the average sensing capabilities of these ground states. Panel (b) displays the QFI distributions for such ground states, beyond their average values. We observe a qualitative change in the shape of the distribution for system sizes beyond the critical size $N^*$, in this case $ N^* =40 $. For $N> N^*$, the probability of finding resourceful states is rapidly suppressed. This behavior is not observed for $N < N^*$. Moreover, as discussed in detail in the End Matter, we find that the skewness increases linearly with $N$ for $N > N^*$, while remaining constant for $N < N^*$. This observation implies that the two regimes separated by $N^*$ are not only a feature of the average value, but also are manifested quantitatively in the underlying distributions. \\

\begin{figure}[t]
    \centering
    \includegraphics[width=0.45\textwidth]{  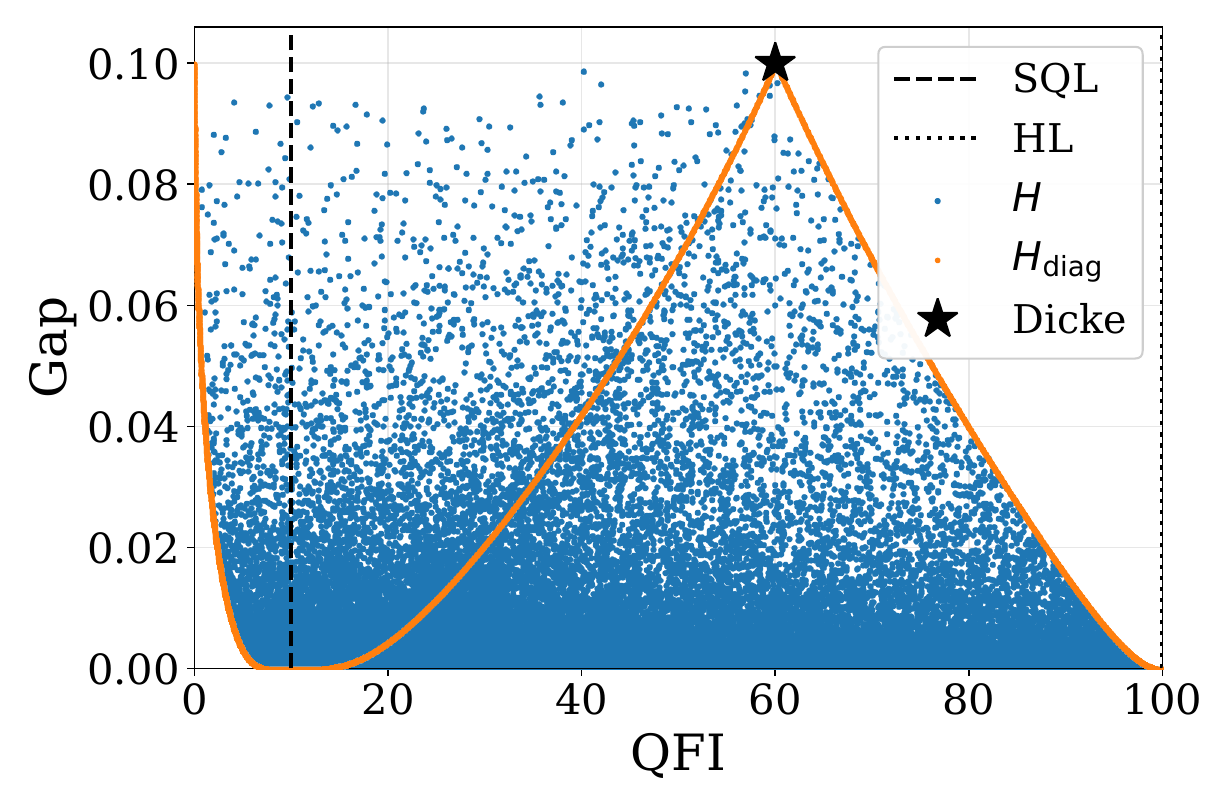}
    \caption{Energy gap and corresponding QFI for $10^5$ ground states of random two-body Hamiltonians for $N=10$ particles and $k=2$. Eq.~\eqref{H_sys} (blue dots). Orange dots correspond to Hamiltonians with only nonzero coefficients, $\Gamma_{200},\Gamma_{020},  \Gamma_{002}$. The star corresponds to  Eq.~\eqref{eq:H_squeezing}, which displays a gap of $12/(N(N+2))$ and a QFI of $N(N/2+1)$. Plots for higher $k$ can be found at the SM.}
    \label{GapVSQFI}
\end{figure}

Next, to further assess the preparation cost of these metrologically useful ground states, we plot the energy gap of the Hamiltonian against the QFI for different random instances with $N=10$ and $k=2$, in Fig.~\eqref{GapVSQFI}. The energy gap, defined as $\lambda_{m+1}(H) - \lambda_{m}(H)$, where $\lambda_m,\lambda_{m+1}$ are the lowest and second-lowest eigenvalues, respectively, can serve as a measure of preparation difficulty. To compare different Hamiltonians, we normalize their energy scales such that $\langle H \rangle_0 = 0, \langle H \rangle_{\infty} = 1$, where $\langle H \rangle_T = \mathrm{Tr}(H e^{-H/T})/\mathrm{Tr}(e^{-H/T})$ is the thermal expectation value. 

 For the QFI above the Dicke state value, in Fig.~\ref{GapVSQFI}, we notice a tradeoff between the energy gap and the corresponding QFI: the highest the QFI, the narrowest is the energy gap of the corresponding Hamiltonian and the ground state becomes more difficult to prepare. 
 The Hamiltonians which maximize the gap for a given QFI (greater than $N(N/2+1)$) of the ground state are of the form $H_{\rm diag} = \Gamma_{200}S_{{200}} + \Gamma_{020}S_{020} + \Gamma_{002}S_{002}$, see orange line in Fig.~\ref{GapVSQFI}. We note that this result is not replicated for $k>2$, see the SM. 
 In order to quantify this tradeoff, we can compute the line connecting the Dicke point $(12/(N(N+2)), N(N/2+1))$ (star in Fig.~\ref{GapVSQFI}) and the Heisenberg point $(0, N^2)$, which upper-bounds the QFI for $N=10$ and $\mathrm{gap}\leqslant 12/(N(N+2)) $. We also verified numerically that the bound holds for greater even $N$ up to $N= 100$. From this observation, we arrive at $\qfi[{\ket{\psi_0}}]\leqslant  N^2 -N^2(N^2-4) \times\mathrm{gap}/24$ for ground states $\ket{\psi_0}$ of Eq.~\eqref{H_sys} and $k=2$. The fact that the QFI is maximized in critical systems, where the gap is closed, was already observed for certain models~\cite{frerot_quantum_2018}, while here we provide a more general bound.

Our approach can be applied to the case in which the generator $G_\theta$ depends on the parameter, as discussed in the SM. In addition, the powerful tools we have developed for calculating Hamiltonians with $k$-body interactions in the symmetric subspace allows us to address many-body metrology \cite{abiuso_fundamental_2025,puig_dynamical_2025} and also nonlinear Hamiltonians, such as interacting encoding Hamiltonians. However, due to the ambiguity in resource accounting \cite{zwierz_general_2010, zwierz_ultimate_2012} and in defining metrological advantage for nonlinear (as opposed to linear) Hamiltonians \cite{imai_metrological_2025}, we do not consider them further in this paper. \\

\textbf{Conclusions}--- In summary, we have studied the metrological usefulness of many-body symmetric ground states of PI Hamiltonians for an even number of particles $N$. As a result, we have found that typical random ground states of few-body PI Hamiltonians with low even-body interactions exhibit Heisenberg-like scaling of the QFI for systems with up to about one hundred (even) spins. We have shown how to sample the couplings so that the energy eigenstates, including the ground state, are Haar-random distributed in the symmetric subspace. Next, we have characterized the distribution of Hamiltonians that lead to such advantage. Motivated by these findings, we have uncovered a tradeoff between the Hamiltonian's gap, which quantifies preparation hardness, and the QFI of the corresponding ground state. As a byproduct of our results, we have developed a dynamical programming algorithm that enables efficient evaluation of spin Hamiltonians in the symmetric subspace.
Our work naturally opens up further research questions. In the context of open quantum systems~\cite{kurdzialek_universal_2025}, it would be beneficial to characterize and analyze the preparation complexity~\cite{marconi_symmetric_2025} of the optimal probe state in more realistic scenarios, taking into consideration robustness against noise and/or particle losses. Another possible avenue for further research is to consider encoding generators beyond one-body interactions. \\

\textbf{Code and data statement}--- Code and data are available from the authors upon reasonable request. \\

\textbf{Acknowledgments}---We thank Jin-Fu Chen for insightful discussions on the computation of $k$-body interactions in the symmetric subspace. 
Majid Hassani and JT acknowledge the support received from the European Union’s Horizon Europe research and innovation programme through the ERC StG FINE-TEA-SQUAD (Grant No. 101040729). This publication is part of the `Quantum Inspire - the Dutch Quantum Computer in the Cloud' project (with project number [NWA.1292.19.194]) of the NWA research program `Research on Routes by Consortia (ORC)', which is funded by the Netherlands Organization for Scientific Research (NWO). GMR acknowledges financial support by the European Union
under ERC Advanced Grant TAtypic, Project No. 101142236.  
MF was supported by the Swiss National Science Foundation Ambizione Grant No. 208886, and by The Branco Weiss Fellowship -- Society in Science, administered by the ETH Z\"{u}rich. Views and opinions expressed are, however, those of the author(s) only and do not necessarily reflect those of the funding institutions. Neither of the funding institutions can be held responsible for them.

\bibliography{references}

{\bf \normalsize 
\begin{center}
    End Matter
\end{center}
}
\section{Haar-random Hamiltonian eigenstates}\label{Haar_random}

Let us begin by defining the concept of vectorization. 
\begin{definition}
Let $A$ be a $d\times d$ Hermitian matrix with elements $\{A_{ij}\}$. We define
$|A\rangle\rangle \in \mathbb{R}^{d^2}$ as
\begin{equation}\label{def_vectorization}
  |A\rangle\rangle \;=\;
  \sqrt{d}\Big(
    \underbrace{A_{ii}}_{i=1~{\mathrm{to}}~d},
    \underbrace{\sqrt{2}\,\mathrm{Re}\,A_{ij}, \sqrt{2}\,\mathrm{Im}\,A_{ij}}_{\mathrm{for}~i<j} \Big)^T. 
\end{equation}
We also define its inverse as $\mathrm{unvec}(|A\rangle\rangle) = A$. 
\end{definition}
We can express the Hamiltonian~\eqref{H_sys} ($d = N+1$) in the vectorized form $|H\rangle \rangle = \sum_{a+b+c\leqslant k} \Gamma_{abc}|S_{abc}\rangle\rangle$. Next, we define $\vec{\Gamma} = (\Gamma_{000}, \Gamma_{001},...,\Gamma_{N00})^T$, where the couplings are sorted in a graded lexicographic order by total degree $k = a+b+c$. The number of interaction terms in $\vec{\Gamma}$ is $\mathbf{N}:=\sum_{k=0}^{N}\binom{k+2}{2}=\frac{(N+3)(N+2)(N+1)}{6}$, thus $\vec{\Gamma}\in \mathbb{R}^{\mathbf{N}}$. In this way, we can express the Hamiltonian as $|H\rangle \rangle = \mathcal{S}\vec{\Gamma}$, where $\mathcal{S} = (|S_{000}\rangle\rangle, |S_{001}\rangle\rangle,...,|S_{N00}\rangle\rangle )$ is a rectangular matrix of size $(N+1)^2\times\mathbf{N}$. Next, we identify the minimal sets of interactions required to construct an arbitrary Hamiltonian. \\

\noindent \textbf{Minimal sets of interactions}--- Since the number of terms, $\mathbf{N}$, exceeds the number of degrees of freedom, $(N+1)^2$, the set $S_{abc}$ is overcomplete. A minimal subset of $(N+1)^2$ interactions can nevertheless be selected to span the space of system Hamiltonians. Starting from the first column in $\mathcal{S}$, we sequentially retain only those columns that increase the rank of the matrix formed by the previously selected columns. Equivalently, a column is retained precisely when it is linearly independent of all previously retained columns. The procedure terminates when the resulting matrix $\mathcal{S}'$ is square and has full rank $(N+1)^2$, and is therefore invertible. Examples of these minimal sets are listed in Table~\ref{tab:bases} of the SM. Note that the minimal sets of interactions are not unique, as the selected interactions depend on the ordering of the interaction terms. In practice, this ordering can be chosen to reflect the increasing experimental effort required to engineer the interactions in the laboratory. Here, we therefore sort the interaction terms by increasing total degree $k=a+b+c$, prioritizing lower-degree interactions. Nonetheless, we observe that such minimal sets must include necessarily full-body interactions, $k = N$. Indeed, we establish the following result: 
\begin{theorem}
   Any minimal set of interactions spanning the space of Hermitian $(N+1)\times(N+1)$ matrices must contain at least one full-body interaction of degree $k=N$.
\end{theorem}
\begin{proof}
The space of Hermitian operators on a spin-$j=N/2$ system decomposes into multipole sectors of ranks $\ell=0,\ldots,N$, with dimensions $2\ell+1$. Interactions $S_{abc}$ of degree $a+b+c=k<N$ contain multipoles only up to rank $N-1$, and therefore span a space of dimension at most $\sum_{\ell=0}^{N-1}(2\ell+1)=N^2<(N+1)^2$. Hence, they cannot span the full space of Hermitian operators, and at least one interaction of degree $k=N$ is necessary.
\end{proof}

Given a minimal set and its corresponding $\mathcal{S}'$, the next theorem determined how to sample the couplings $\Gamma$ in order to produce a Hamiltonian with Haar-random eigenstates.

\begin{theorem}
Let $\mathcal{S}'$ be an invertible matrix whose columns are the vectorized interactions in a minimal set and \(\vec{\Gamma}'\in\mathbb{R}^{(N+1)^2}\) the  corresponding coupling vector. If $\vec{\Gamma}'$ is sampled from a multivariate normal distribution of mean $\mathbf{0}$ and covariance $\mathcal{S}'^{-1}(\mathcal{S}'^{-1})^{T}$, $\vec{\Gamma}'\sim\mathcal{N}(\mathbf{0},\mathcal{S}'^{-1}(\mathcal{S}'^{-1})^{T})$, then the eigenstates of the Hamiltonain $H=\mathrm{unvec}(\mathcal{S}'\vec{\Gamma}')$ are Haar-random.  
\end{theorem}
\begin{proof}
A Hamiltonian drawn from the Gaussian unitary ensemble (GUE) has eigenvectors distributed according to the Haar measure \cite{mehta_random_2004,anderson_introduction_2010}. Under the real vectorization convention defined above, a suitably normalized GUE Hamiltonian satisfies $|H\rangle\rangle\sim
\mathcal{N}\left(\mathbf{0},\mathbb{I}\right)$. Since $|H\rangle\rangle=\mathcal{S}'\vec{\Gamma}'$, we have
$\vec{\Gamma}'=\mathcal{S}'^{-1}|H\rangle\rangle$. By the linear transformation theorem for multivariate normal distributions, if $(X\sim\mathcal{N}(\mu,\Sigma))$, then $(AX\sim\mathcal{N}(A\mu,A\Sigma A^T))$. Consequently, $\vec{\Gamma}'\sim\mathcal{N}\left(
\mathbf{0},
\mathcal{S}'^{-1}(\mathcal{S}'^{-1})^T
\right)$. Thus, sampling the coupling vector from the stated distribution makes $H$, GUE distributed and its eigenvectors are therefore Haar distributed.
\end{proof}

We have also verified the sampling procedure numerically. 
For $N=10$, sampling $\vec{\Gamma}$ from this distribution and diagonalizing the resulting $H$, the ground-state QFI averaged over $5\times10^{4}$ samples is $36.64\pm0.04$, in agreement with the Haar average value $N(N+1)/3\approx36.67$ (the same holds for every eigenvector index and for $N=4,6,8,12$).

\section{Variance and skewness of the QFI distribution}

\begin{figure*}[h]
    \centering
  {\includegraphics[width=0.8\linewidth]{  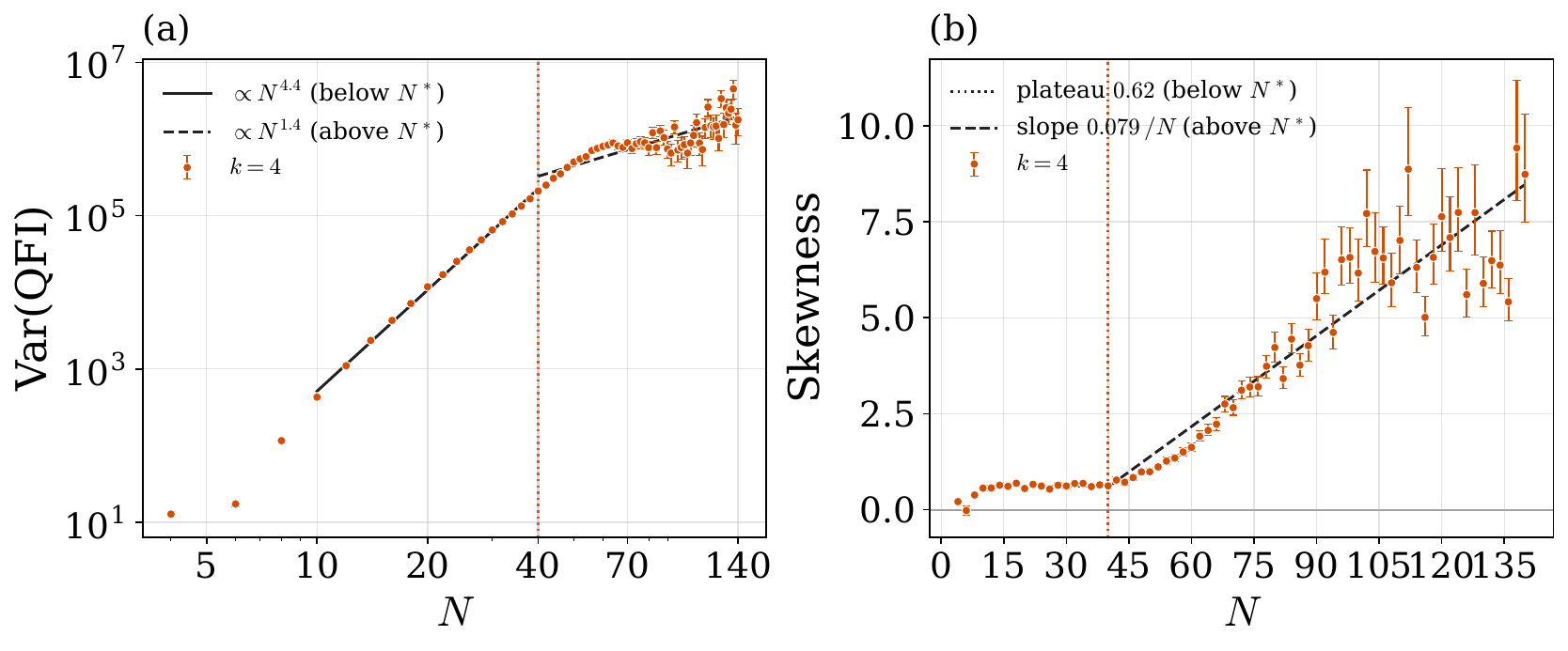}} \\
    {\includegraphics[width=0.8\linewidth]{  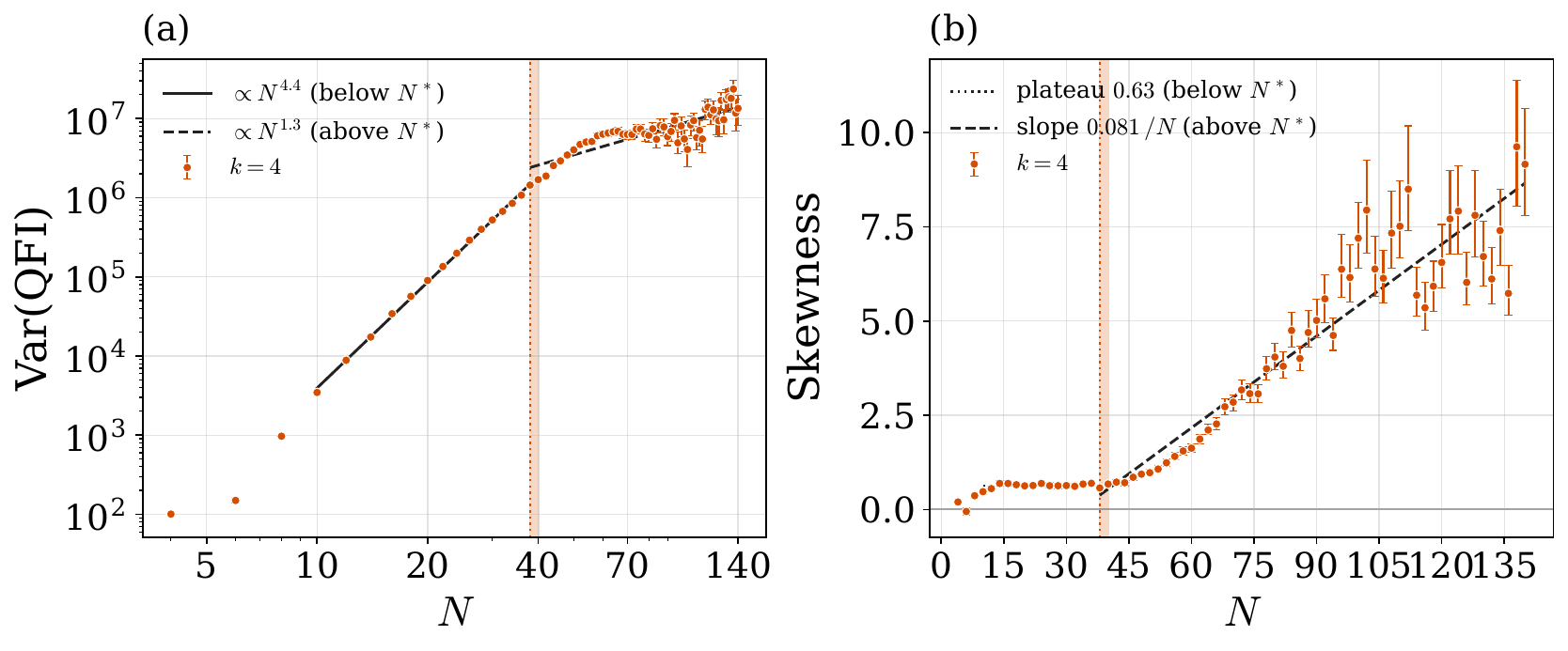}}
    \caption{Variance (left) and skewness (right) of the QFI distribution at $k=4$ as functions of $N$ (even $N$, 2000 realizations per $(N,k)$, error bars are bootstrap 95\% confidence intervals) for (a) the parameter-independent encoding of Fig.~\ref{H_X-random_sampling} ($G_\theta =\theta \sigma_x /2$), (b) the parameter-dependent encoding $G_\theta = \cos(\theta)\sigma _{z}+\sin(\theta)\sigma _{x}$ (see the SM). The dotted line marks $N^{*}(k{=}4)$, read off the mean as in the inset of Fig.~\ref{H_X-random_sampling}(a). Black lines are fits below and above $N^{*}$.}
    \label{fig:moments}
\end{figure*}

Figure~\ref{fig:moments} shows the variance and the skewness of the QFI distribution of Fig.~\ref{H_X-random_sampling} for $k=4$ as functions of $N$. Both moments display two regimes separated by the same $N^{*}$ obtained from the mean (dotted line). Below $N^{*}$ the variance grows as $N^{4.4}$, i.e. as fast as the squared mean, so that the relative variance stays at $0.5$--$0.7$ for $N=20$--$49$, and the skewness sits at a plateau $\simeq0.6$. The shape of the distribution does not depend on $N$.
Above $N^{*}$ the variance grows only as $N^{1.4}$ while the mean detaches from $N(N+1)/3$, so that the relative variance increases with $N$, and the skewness rises linearly with a slope of $0.08$ per spin, reaching $\approx4$ at $N=80$ and $\approx6$ at $N=100$.
The parameter-dependent generator (see the SM) yields the same picture, Fig.~\ref{fig:moments} (b), $N^{*}=38$, plateau $0.6$, slope $0.08$ per spin.

\clearpage
\onecolumngrid
{\bf \normalsize 
\begin{center}
    SUPPLEMENTARY MATERIAL
\end{center}
}
\section{Upper bound on the QFI---Derivation of Eq.~(\ref{QFI_PDH_Variance_upper_bound})}\label{Appendix_upper_bound}
Here we derive an upper bound on the QFI of unitary evolution and a pure probe state. The approach is nearly the same as the one used in Ref. \cite{boixo_generalized_2007}. Let us begin by recalling the closed-form formula of the QFI
\begin{equation}\label{QFI_PDH_der_upp_b1}
    \qfi[\ket{\psi_\theta}]= 4\Delta ^{2} \mathbf{K}_{\theta}=4(\average{\mathbf{K}_{\theta}^{2}}-\average{\mathbf{K}_{\theta}}^{2}).
\end{equation}
To provide an upper bound on the variance in Eq.~(\ref{QFI_PDH_der_upp_b1}), one can exploit the operator seminorm with following properties
\begin{itemize}
    \item (\textbf{P1}): for Hermitian operator $\mathpzc{A}$ is defined as $\lambda_{M}(\mathpzc{A})-\lambda _{m}(\mathpzc{A})$,
    \item (\textbf{P2}): it is called seminorm (not a norm) because for $\mathpzc{A}=\alpha\mathbb{1}$, $\Vert\mathpzc{A}\Vert_{\text{sn}}=0$, but $\mathpzc{A}\neq 0$, unless $\alpha\neq 0$,
    \item (\textbf{P3}): \textbf{triangle inequality (commuting operators)}: If $[\mathpzc{A},\mathpzc{B}] = 0$, then 
    $\Vert\mathpzc{A}+\mathpzc{B}\Vert _{\text{sn}}=\Vert\mathpzc{A}\Vert _{\text{sn}}+\Vert\mathpzc{B}\Vert _{\text{sn}}$, 
    \item (\textbf{P4}): \textbf{unitary invariance}: $\Vert U\mathpzc{A}U^{\dagger}\Vert_{\text{sn}}=\Vert \mathpzc{A}\Vert_{\text{sn}}$ for any unitary $U$.
\end{itemize}
Since $\mathbf{K}_{\theta}$ is a Hermitian operator, Eq.~(\ref{QFI_PDH_der_upp_b1}) yields
\begin{equation}\label{QFI_PDH_der_upp_b2}
    \qfi[\ket{\psi_\theta}]\leqslant\Vert\mathbf{K}_{\theta}\Vert _{\text{sn}}^{2}.
\end{equation}
In order to upper bound Eq.~(\ref{QFI_PDH_der_upp_b2}), we start from the definition of $\mathbf{K}_{\theta}$
\begin{align}
    \Vert\mathbf{K}_{\theta}\Vert _{\text{sn}}&~=\Big\Vert i\partial_{\theta}\mathbf{U}_{\theta}~\mathbf{U}_{\theta}^{\dagger}\Big\Vert _{\text{sn}}\nonumber\\
    &~=\Big\Vert i\partial_{\theta}U_{\theta}^{\otimes N}~U_{\theta}^{\otimes N\dagger}\Big\Vert _{\text{sn}}\nonumber\\
    &~=\Big\Vert i\big (\sum _{i=1}^{N}U_{\theta}\otimes U_{\theta}\otimes\cdots\otimes \underbrace{\partial _{\theta}U_{\theta}}_{i^{\text{th}}}\otimes\cdots U_{\theta}\big)~U_{\theta}^{\otimes N\dagger}\Big\Vert _{\text{sn}}\nonumber\\
    &~=\Big\Vert \sum _{i=1}^{N}\mathbb{1}\otimes \mathbb{1}\otimes\cdots\otimes \underbrace{i\partial _{\theta}U_{\theta}~U_{\theta}^{\dagger}}_{i^{\text{th}}}\otimes\cdots \mathbb{1}\Big\Vert _{\text{sn}}\nonumber\\
    &\overset{(\textbf{P3})}{=}\sum _{i=1}^{N}\Big\Vert \mathbb{1}\otimes \mathbb{1}\otimes\cdots\otimes K_{\theta}\otimes\cdots \mathbb{1}\Big\Vert _{\text{sn}}\nonumber\\
    &=N\Vert K_{\theta}\Vert _{\text{sn}}.\nonumber\\
\end{align}
From whence
\begin{equation} \label{bound}
    \qfi[\ket{\psi_\theta}]=4\Delta ^{2} \mathbf{K}_{\theta}\leqslant\Vert\mathbf{K}_{\theta}\Vert _{\text{sn}}^{2} = N^{2}\Vert K _{\theta}\Vert_{\text{sn}}^{2}.
\end{equation}
\section{Preliminaries on the symmetric subspace}\label{prliminary_subspace}
This appendix presents essential background on the symmetric subspace that is necessary for deriving subsequent results. The symmetric subspace can be expanded by the so-called Dicke states. In the computational basis, an $N$-qubit Dicke state with $n$ excitations is given by 
\begin{equation}\label{dicke-state-def}
    \ket{D_{N}^{n}}=\frac{1}{\sqrt{\binom{N}{n}}}\sum_{\mathbf{i}}\delta(\vert\mathbf{i}\vert-n)\ket{\mathbf{i}},
\end{equation}
where $\mathbf{i}=i_{0}i_{1}\cdots i_{N-1}$ in which $i_{\gamma}\in\{0,1\}$ and $\vert\mathbf{i}\vert$ denotes the number of ones in the binary representation. Any symmetric quantum state, $\ket{\psi}$, can be written in the symmetric subspace as follows
\begin{equation}\label{state-sym-sub}
    \ket{\psi}=\sum _{n=0}^{N}C_{n}\ket{D_{N}^{n}}.
\end{equation}
where $C_{n}\in\mathbb{C}$. To reduce the complexity of the problem from $2^{N}$ in the computational basis to $(N+1)$ within the symmetric subspace (spanned by the Dicke states defined in Eq.~(\ref{dicke-state-def})), one can define the following projector operator
\begin{equation}\label{projec-symm}
    \mathpzc{P}=\mathpzc{V}^{\dagger}\mathpzc{V},
\end{equation}
where
\begin{equation}\label{def-V}
    \mathpzc{V}=\sum _{n=0}^{N}\ket{n}\bra{D_{N}^{n}},
\end{equation}
in which $\ket{n}$ and $\ket{D_{N}^{n}}$ are in the symmetric subspace ($(N+1)$-dimension) and the computational space ($2^{N}$-dimension), respectively, and satisfies $\mathpzc{V}\mathpzc{V}^{\dagger}=\mathbb{1}_{(N+1)\times(N+1)}$.

An alternative definition of the symmetric subspace can be given in terms of permutation operators \cite{mele_introduction_2024}. The symmetric group $\mathcal{S}_{N}$ consisting of permutations of $N$ elements, where 
\begin{equation}\label{permutation_elements}
    \sigma: \{1,2, \cdots, N\}\rightarrow\{1,2, \cdots, N\}, ~~~~~~~~~~~~\forall\sigma\in \mathcal{S}_{N}.
\end{equation}
One can define the permutation matrix $U_{\sigma}$ which satisfies
\begin{equation}\label{permut_psi}
    U_{\sigma}(\ket{\psi _{1}}\otimes\ket{\psi_{2}}\otimes\cdots\ket{\psi_{N}})=\ket{\psi_{\sigma^{-1}(1)}}\otimes\ket{\psi_{\sigma^{-1}(2)}}\otimes\cdots\ket{\psi_{\sigma^{-1}(N)}}.
\end{equation} 
Specifically, let us define the symmetric subspace as the space that remains invariant under the action of any permutation matrix $U_{\sigma}$ as 
\begin{equation}\label{symmetric_permut_def}
    \text{Sym}:=\{\ket{\psi}\vert~~~U_{\sigma}\ket{\psi}=\ket{\psi}, \forall\sigma\in\mathcal{S}_{N}\}.
\end{equation}
Hence, one can define an alternative definition of the projector operator onto the symmetric subspace, Eq.~(\ref{projec-symm}), as follows
\begin{equation}\label{projec_symm_permutation}
    \mathpzc{P}=\frac{1}{N!}\sum_{\sigma\in\mathcal{S}_{N}}U_{\sigma}.
\end{equation}
It is straightforward to check that Eq.~(\ref{projec_symm_permutation}) satisfies all the conditions of projector operation.
\section{Computing the QFI for the unitary evolution in the symmetric subspace}\label{Com_QFI_symm_sub}
Here, we derive a general formula for calculating the QFI in the case of unitary evolution in the symmetric subspace. Computing the QFI in the symmetric subspace simplifies calculations and allows us to consider larger $N$. Consider the evolution where the unknown parameter $\theta$ is encoded via local unitary evolution $U_{\theta}=\re^{-iG_{\theta}}$ in which $G_{\theta}$ is a Hermitian operator. Hence, the sampling operator can be defined as 
\begin{equation}\label{gen-Uni-evo}
    \mathbf{U}_{\theta}=U_{\theta}^{\otimes N}.
\end{equation}
It is straightforward to check that for any $2\times 2$ unitary evolution like
\begin{align}\label{gen-Uni_2}
    U_{\theta}&=\left(\begin{array}{cc}
    \mathcal{U}_{00}(\theta)&\mathcal{U}_{01}(\theta)\\ 
    \mathcal{U}_{10}(\theta)&\mathcal{U}_{11}(\theta)\\
    \end{array}\right),
\end{align}

the representation of the sampling operator, $\mathbf{U}_{\theta}$ (Eq.~(\ref{gen-Uni-evo})), in the computational basis is given by 
\begin{align} \label{gen-Uni-evo-comp-basis}
    \mathbf{U}_{\theta} = \sum_{\mathbf{i}, \mathbf{j} \in \{ 0,1\}^N} \mathcal{U}_{i_0,j_0}(\theta)~\mathcal{U}_{i_1,j_1}(\theta)~...~\mathcal{U}_{i_{N-1},j_{N-1}}(\theta)  \ket{i_0, i_1,...,i_{N-1}}\bra{j_0,j_1,...,j_{N-1}} 
\end{align}
where, as defined above, $\mathbf{i}=i_{0}i_{1}\cdots i_{N-1}$ and $\mathbf{j}=j_{0}j_{1}\cdots j_{N-1}$.
\begin{theorem}\label{T1:U_commut_P}
The sampling operator $\mathbf{U}_{\theta}$ commutes with the projector operator onto the symmetric subspace, $\mathpzc{P}$,
\begin{equation}\label{U_commut_P}
    [\mathbf{U}_{\theta}, \mathpzc{P}]=0.
\end{equation}
\end{theorem}
\textit{Proof:} Given the explicit form of $\mathpzc{P}$ in Eq.~(\ref{projec_symm_permutation}), it is necessary to examine the commutation relation between $\mathbf{U}_{\theta}$ and $U_{\sigma}$.
\begin{lemma}\label{L1:U_commut_sigma}
The sampling operator $\mathbf{U}_{\theta}$ commutes with permutation matrix $U_{\sigma},~~~\forall\sigma\in\mathcal{S}_{N}$ (Eq.~(\ref{permut_psi})).
\end{lemma}
\textit{Proof:} Let consider
\begin{align}\label{U_commut_sigma_proof}
    U_{\sigma}\mathbf{U}_{\theta}(\ket{\psi _{1}}\otimes\ket{\psi_{2}}\otimes\cdots\ket{\psi_{N}})&\overset{(\ref{gen-Uni-evo})}{=}U_{\sigma}U_{\theta}^{\otimes N}(\ket{\psi _{1}}\otimes\ket{\psi_{2}}\otimes\cdots\otimes\ket{\psi_{N}})\nonumber\\
    &\,~=U_{\sigma}(U_{\theta}\ket{\psi _{1}}\otimes U_{\theta}\ket{\psi_{2}}\otimes\cdots\otimes U_{\theta}\ket{\psi_{N}})\nonumber\\
    &\overset{(\ref{permut_psi})}{=}\ket{\psi _{\sigma^{-1}(1)}(\theta)}\otimes \ket{\psi_{\sigma^{-1}(2)}(\theta)}\otimes\cdots\otimes \ket{\psi_{\sigma^{-1}(N)}(\theta)}\nonumber\\
    &\,~=U_{\theta}\ket{\psi _{\sigma^{-1}(1)}}\otimes U_{\theta}\ket{\psi_{\sigma^{-1}(2)}}\otimes\cdots\otimes U_{\theta}\ket{\psi_{\sigma^{-1}(N)}}\nonumber\\
    &\,~=\mathbf{U}_{\theta}U_{\sigma}(\ket{\psi_{1}}\otimes\ket{\psi_{2}}\otimes\cdots\otimes\ket{\psi_{N}}).
\end{align}
Since Eq.~(\ref{U_commut_sigma_proof}) holds for any arbitrary $\ket{\psi _{1}}, \ket{\psi_{2}}, \cdots ,\ket{\psi_{N}}$, then
\begin{equation}\label{U_commut_sigma}
    [\mathbf{U}_{\theta}, U_{\sigma}] =0,~~~~~~~~~~~~~~~~~~~\forall\sigma\in\mathcal{S}_{N}.
\end{equation}
$\hfill\blacksquare$\\
Therefore, Eqs. (\ref{projec_symm_permutation}) and (\ref{U_commut_sigma}) complete the proof.

$\hfill\blacksquare$\\
\textbf{Corollary 1} Since $[\mathbf{U}_{\theta}, \mathpzc{P}]=0$, then
\begin{equation}\label{1st_deriv_U_commut_P}
   [\partial_{\theta}\mathbf{U}_{\theta}, \mathpzc{P}]=0.
\end{equation}
To compute the QFI, Eq.~(\ref{QFI-pure-states}), let us consider the case in which the sampling operator is given by Eq.~(\ref{gen-Uni-evo}) and the initial state $\ket{\psi_{0}}$ is chosen from the symmetric subspace. Hence
\begin{equation}\label{psi_after_evolution}
    \ket{\psi _{\theta}}=\mathbf{U}_{\theta}\ket{\psi _{0}},
\end{equation}
and
\begin{equation}\label{invariant_symm_subspace}
    \ket{\psi_{0}}=\mathpzc{P}\ket{\psi_{0}}.
\end{equation}
Whence
\begin{align}
    \ket{\psi_{\theta}}&=\mathbf{U}_{\theta}\mathpzc{P}\ket{\psi _{0}},\label{psi_after_evolution_symm}\\
    \ket{\partial_{\theta}\psi_{\theta}}&=\partial_{\theta}\mathbf{U}_{\theta}\mathpzc{P}\ket{\psi _{0}}.\label{psi_dot_after_evolution_symm}
\end{align}
Substituting Eqs. (\ref{psi_after_evolution_symm},\ref{psi_dot_after_evolution_symm}) in the QFI relation, Eq.~(\ref{QFI-pure-states}), yields to the QFI of the pure state in the symmetric subspace
\begin{align}\label{QFI-symmetric-subspace-dev1}
  \qfi^{(\text{sym.})}[\ket{\psi_{\theta}}]&~~~~~~~= ~~~~~4\Big(\average{\psi_{0}\vert\mathpzc{P}\partial_{\theta}\mathbf{U}^{\dagger}_{\theta}~\partial_{\theta}\mathbf{U}_{\theta}\mathpzc{P}\vert\psi_{0}}-\average{\psi_{0}\vert\mathpzc{P}\mathbf{U}^{\dagger}_{\theta}~\partial_{\theta}\mathbf{U}_{\theta}\mathpzc{P}\vert\psi_{0}}\average{\psi_{0}\vert\mathpzc{P}\partial_{\theta}\mathbf{U}^{\dagger}_{\theta}~\mathbf{U}_{\theta}\mathpzc{P}\vert\psi_{0}}\Big)\nonumber\\
  &\,~~\overset{(\mathpzc{P}^{2}=\mathpzc{P})}{=}~4\Big(\average{\psi_{0}\vert\mathpzc{P}~\mathpzc{P}\partial_{\theta}\mathbf{U}^{\dagger}_{\theta}~\partial_{\theta}\mathbf{U}_{\theta}\mathpzc{P}~\mathpzc{P}\vert\psi_{0}}\Big)\nonumber\\
  &\,~~~~~~~~~~~\,-4\Big(\average{\psi_{0}\vert\mathpzc{P}~\mathpzc{P}\mathbf{U}^{\dagger}_{\theta}~\partial_{\theta}\mathbf{U}_{\theta}\mathpzc{P}~\mathpzc{P}\vert\psi_{0}}\average{\psi_{0}\vert\mathpzc{P}~\mathpzc{P}\partial_{\theta}\mathbf{U}^{\dagger}_{\theta}~\mathbf{U}_{\theta}\mathpzc{P}~\mathpzc{P}\vert\psi_{0}}\Big)\nonumber\\
    &~~\overset{(\ref{U_commut_P}),(\ref{1st_deriv_U_commut_P})}{=}4\Big(\average{\psi_{0}\vert\mathpzc{P}\partial_{\theta}\mathbf{U}^{\dagger}_{\theta}\mathpzc{P}\partial_{\theta}\mathbf{U}_{\theta}\mathpzc{P}\vert\psi_{0}}\Big)\nonumber\\
    &\,~~~~~~~~~~\,-4\Big(\average{\psi_{0}\vert\mathpzc{P}\mathbf{U}^{\dagger}_{\theta}\mathpzc{P}\partial_{\theta}\mathbf{U}_{\theta}\mathpzc{P}\vert\psi_{0}}\average{\psi_{0}\vert\mathpzc{P}\partial_{\theta}\mathbf{U}^{\dagger}_{\theta}\mathpzc{P}\mathbf{U}_{\theta}\mathpzc{P}\vert\psi_{0}}\Big)\nonumber\\
    &\,~~~~~\overset{(\ref{projec-symm})}{=}~~~4\Big(\average{\phi_{0}\vert\partial_{\theta}~\mathbf{W}^{\dagger}_{\theta}\partial_{\theta}\mathbf{W}_{\theta}\vert\phi_{0}}-\average{\phi_{0}\vert\mathbf{W}^{\dagger}_{\theta}~\partial_{\theta}\mathbf{W}_{\theta}\vert\phi_{0}}\average{\phi_{0}\vert\partial_{\theta}\mathbf{W}^{\dagger}_{\theta}~\mathbf{W}_{\theta}\vert\phi_{0}}\Big),
\end{align}
where
\begin{align}
    \ket{\phi_{0}}&:=\mathpzc{V}\ket{\psi _{0}}\label{def_phi_0},\\
    \mathbf{W}_{\theta}&:=\mathpzc{V}~\mathbf{U}_{\theta}~\mathpzc{V}^{\dagger}\label{def_W}.
\end{align}
As explicitly shown by Eq.~(\ref{def_phi_0}), $\ket{\phi _{0}}$ belongs to the $(N+1)$-dimensional symmetric subspace. Hence, it can be written in the following form
\begin{equation}\label{phi_0_symm_sub}
    \ket{\phi _{0}}=\sum _{n=0}^{N}\alpha _{n}\ket{n},
\end{equation}
where $\alpha_{n}\in\mathbb{C}$ and $\sum _{n}\vert\alpha _{n}\vert ^{2}=1$. Crucially, calculating the QFI within the symmetric subspace, Eq.~(\ref{QFI-symmetric-subspace-dev1}), requires computing $\mathbf{W}_{\theta}$, Eq.~(\ref{def_W}). Indeed, this operator denotes the projection of the sampling operator, $\mathbf{U}_{\theta}$, from the $2^{N}$-dimensional computational basis onto the $(N+1)$-dimensional symmetric subspace. The explicit derivation of $\mathbf{W}_{\theta}$ is as follows
\begin{align}\label{deriv_W_1}
    \mathbf{W}_{\theta}&=\mathpzc{V}~\mathbf{U}_{\theta}~\mathpzc{V}^{\dagger}\nonumber\\
    &\overset{(\ref{def-V})}{=}\sum _{n,m=0}^{N}\mathpzc{C}_{mn}\ket{m}\bra{n},
\end{align}
where
\begin{equation}\label{C_coeff_def}
    \mathpzc{C}_{mn}:=\bra{D_{N}^{m}}\mathbf{U}_{\theta}\ket{D_{N}^{n}}.
\end{equation}
Substituting Eqs. (\ref{dicke-state-def},\ref{gen-Uni-evo-comp-basis}) in Eq.~(\ref{C_coeff_def}), gives
\begin{align}\label{Coeff_elementary}
    \mathpzc{C}_{mn}&=\frac{1}{\sqrt{\binom{N}{m}\binom{N}{n}}}\sum_{\mathbf{i},\mathbf{j}}\delta(\vert\mathbf{i}\vert-m)\delta(\vert\mathbf{j}\vert-n)\prod_{\gamma = 0}^{ N-1}\mathcal{U}_{i_{\gamma},j_{\gamma}}(\theta).
\end{align}
In order to compute $\mathpzc{C}_{mn}$, we define four vectors $\omega _{i}$ for $i\in\{0, 1, 2, 3\}$
\begin{align}
    \omega _{0}&=\left(\begin{array}{cc}
                0\\ 
                0\\
                \end{array}\right),\label{w0}\\
    \omega _{1}&=\left(\begin{array}{cc}
                0\\ 
                1\\
                \end{array}\right),\label{w1}\\
    \omega _{2}&=\left(\begin{array}{cc}
                1\\ 
                0\\
                \end{array}\right),\label{w2}\\
    \omega _{3}&=\left(\begin{array}{cc}
                1\\ 
                1\\
                \end{array}\right).\label{w3}
\end{align}
The first and second entries of $\omega _{i}$ correspond to elements from strings $\mathbf{i}$ and $\mathbf{j}$ respectively. These vectors are used to determine $\vert\mathbf{i}\vert$ and $\vert\mathbf{j}\vert$ and to select the appropriate $\mathcal{U}_{i_{\gamma},j_{\gamma}}$. Therefore, one can obtain
\begin{align}
   \vert\mathbf{i}\vert&=\bar{\omega}_{2}+\bar{\omega} _{3},\label{norm_i}\\
   \vert\mathbf{j}\vert&=\bar{\omega} _{1}+\bar{\omega} _{3},\label{norm_j}
\end{align}
where $\bar{\omega} _{i}$ counts the number of times where $\omega _{i}$ appears in the $\mathbf{i}$ and $\mathbf{j}$ strings. For example in the case where 
\begin{align*}
    \mathbf{i}&=~~\,0~~~~~~\,0~~~~~~\,1~~~~~~\,0~~~~~~~\,1~~,\nonumber\\
    \mathbf{j}&=\underbrace{0}_{\omega _{0}}~\underbrace{0}_{\omega _{0}}~\underbrace{0}_{\omega _{2}}~\underbrace{1}_{\omega _{1}}~\underbrace{1}_{\omega _{3}},
\end{align*}
$\bar{\omega}_{0}=2$, $\bar{\omega}_{1}=1$, $\bar{\omega}_{2}=1$, and $\bar{\omega}_{3}=1$. Then $\vec{\omega}:=(\bar{\omega}_{0},\bar{\omega}_{1},\bar{\omega}_{2},\bar{\omega}_{3})=(2,1,1,1)$.

Since $N=\sum_{i=0}^{3}\bar{\omega} _{i}$, one needs to count the number of different partitions of $N$ where the $\{\omega _{i}\}$'s are the parts by using
\begin{equation}\label{combinotorial_w}
    \binom{N}{\bar{\omega} _{0}}\binom{N-\bar{\omega} _{0}}{\bar{\omega} _{1}}\binom{N-\bar{\omega} _{0}-\bar{\omega} _{1}}{\bar{\omega} _{2}}\binom{N-\bar{\omega} _{0}-\bar{\omega} _{1}-\bar{\omega} _{2}}{\bar{\omega} _{3}}=\frac{N!}{\bar{\omega} _{0}!\bar{\omega} _{1}!\bar{\omega} _{2}!\bar{\omega} _{3}!}.
\end{equation}
Applying these consideration, Eq.~(\ref{Coeff_elementary}) yields
\begin{equation}\label{C_coeff_final}
    \mathpzc{C}_{mn}=\sum_{\vec{\omega}~\vdash N}\frac{N!}{\bar{\omega} _{0}!\bar{\omega} _{1}!\bar{\omega} _{2}!\bar{\omega} _{3}!}\frac{\delta(\bar{\omega} _{2}+\bar{\omega} _{3}-m)\delta(\bar{\omega} _{1}+\bar{\omega} _{3}-n)}{\sqrt{\binom{N}{m}\binom{N}{n}}}\mathcal{U}_{00}^{\bar{\omega} _{0}}(\theta)\mathcal{U}_{01}^{\bar{\omega} _{1}}(\theta)\mathcal{U}_{10}^{\bar{\omega} _{2}}(\theta)\mathcal{U}_{11}^{\bar{\omega} _{3}}(\theta),
\end{equation}
and its first derivative with respect to the unknown parameter is given by
\begin{align}\label{partial_C_coeff_final}
    \partial_{\theta}\mathpzc{C}_{mn}&=\sum_{\vec{\omega}~\vdash N}\frac{N!}{\bar{\omega} _{0}!\bar{\omega} _{1}!\bar{\omega} _{2}!\bar{\omega} _{3}!}\frac{\delta(\bar{\omega} _{2}+\bar{\omega} _{3}-m)\delta(\bar{\omega} _{1}+\bar{\omega} _{3}-n)}{\sqrt{\binom{N}{m}\binom{N}{n}}}\nonumber\\
    &~~~~~~~~~~~~~~~~~~~~~\Big(\bar{\omega} _{0}\partial _{\theta}\mathcal{U}_{00}(\theta)\mathcal{U}_{00}^{\bar{\omega} _{0}-1}(\theta)\mathcal{U}_{01}^{\bar{\omega} _{1}}(\theta)\mathcal{U}_{10}^{\bar{\omega} _{2}}(\theta)\mathcal{U}_{11}^{\bar{\omega} _{3}}(\theta)+\bar{\omega} _{1}\partial _{\theta}\mathcal{U}_{01}(\theta)\mathcal{U}_{00}^{\bar{\omega} _{0}}(\theta)\mathcal{U}_{01}^{\bar{\omega} _{1}-1}(\theta)\mathcal{U}_{10}^{\bar{\omega} _{2}}(\theta)\mathcal{U}_{11}^{\bar{\omega} _{3}}(\theta)\nonumber\\
    &~~~~~~~~~~~~~~~~~~~~~~~\bar{\omega} _{2}\partial _{\theta}\mathcal{U}_{10}(\theta)\mathcal{U}_{00}^{\bar{\omega} _{0}}(\theta)\mathcal{U}_{01}^{\bar{\omega} _{1}}(\theta)\mathcal{U}_{10}^{\bar{\omega} _{2}-1}(\theta)\mathcal{U}_{11}^{\bar{\omega} _{3}}(\theta)+\bar{\omega} _{3}\partial _{\theta}\mathcal{U}_{11}(\theta)\mathcal{U}_{00}^{\bar{\omega} _{0}}(\theta)\mathcal{U}_{01}^{\bar{\omega} _{1}}(\theta)\mathcal{U}_{10}^{\bar{\omega} _{2}}(\theta)\mathcal{U}_{11}^{\bar{\omega} _{3}-1}(\theta)\Big)
\end{align}
where $ \vec{\omega}\vdash N $ indicates that $N$ is partitioned as $\vec{\omega}$. By exploiting Eqs. (\ref{phi_0_symm_sub},\ref{deriv_W_1},\ref{C_coeff_final},\ref{partial_C_coeff_final}), Eq.~(\ref{QFI-symmetric-subspace-dev1}) can be cast in this form
\begin{equation}\label{QFI-symmetric-subspace}
    \qfi^{(\text{sym.})}[\ket{\psi_{\theta}}]=4\Big[ \sum_{m,n,l=0}^{N}\alpha _{m}^{*}\alpha _{n}(\partial _{\theta}\mathpzc{C}_{lm})^{*}~\partial _{\theta}\mathpzc{C}_{ln}-\Big\vert \sum_{m,n,l=0}^{N}\alpha _{m}^{*}\alpha _{n}\mathpzc{C}_{lm}^{*}~\partial _{\theta}\mathpzc{C}_{ln}\Big\vert ^{2}\Big].
\end{equation}
\section{Computing the PI $k$-body interaction terms in the symmetric subspace}
\label{app:PIk-boday_corre}

We present a recursive formula for computing the PI $k$-body interaction terms in the symmetric subspace.
Let us first introduce the one-body interaction terms as follows
\begin{equation}\label{one-body-cor}
    S_{\mu}=\sum_{i=1}^{N}\frac{1}{2}\sigma _{\mu}^{(i)},
\end{equation}

where $N$ is the number of sites (observers in the multipartite Bell experiment) and $\frac{1}{2}\sigma _{\mu}^{(i)}$ ($\mu\in \{x, y, z\}$) represents the tensor product of the Pauli matrix $\frac{1}{2}\sigma _{\mu}$ acting at the $i$-th site, with identity matrices elsewhere ($\frac{1}{2}\sigma_\mu^{(i)} = \mathbb{1}^{i-1} \otimes \frac{1}{2}\sigma_\mu \otimes \mathbb{1}^{N-i-1}$). It is straightforward to check that the one-body interaction terms satisfy the following commutation relations
\begin{equation}\label{commute-rel}
    [S_{x}, S_{y}]=iS_{z}, ~~~~~~~~~~~~~~~~[S_{y}, S_{z}]=iS_{x}, ~~~~~~~~~~~~~~~~ [S_{z}, S_{x}]=iS_{y}.
\end{equation}
Equation (\ref{one-body-cor}) can be generalized to define the $k$-body interaction terms as 
\begin{equation}\label{k-body-cor}
    S_{\mu_{1}\mu_{2}\cdots\mu_{k}}=\frac{1}{k!} \mathlarger{\mathlarger{\sum}}_{i_{1}\neq i_{2}\neq\cdots\neq i_{k}=1}^{N}\frac{1}{2}\sigma _{\mu_{1}}^{(i_{1})}\,.\,\frac{1}{2}\sigma _{\mu_{2}}^{(i_{2})}\,.\,\cdots\, .\,\frac{1}{2}\sigma _{\mu_{k}}^{(i_{k})}.
\end{equation}
In order to find a recursive relation to computing $k$-body interaction terms from lower-order interaction terms, let us begin from 
\begin{align}\label{deriv-k_body-p1}
    &S_{\mu_{1}\mu_{2}\cdots\mu_{k-1}}\,S_{\mu_{k}}=\nonumber\\
&\Bigg(\mathlarger{\mathlarger{\sum}}_{i_{1}\neq i_{2}\neq\cdots\neq i_{k-1}=1}^{N}\frac{1}{2}\sigma _{\mu_{1}}^{(i_{1})}\,.\,\frac{1}{2}\sigma _{\mu_{2}}^{(i_{2})}\,.\,\cdots\,.\,\frac{1}{2}\sigma _{\mu_{k-1}}^{(i_{k-1})}\Bigg) \Bigg(\sum_{i_{k}=1}^{N}\frac{1}{2}\sigma _{\mu_{k}}^{(i_{k})}\Bigg) =\nonumber\\
    &\mathlarger{\mathlarger{\sum}}_{i_{1}\neq i_{2}\neq\cdots\neq i_{k}=1}^{N}\frac{1}{2}\sigma _{\mu_{1}}^{(i_{1})}\,.\,\frac{1}{2}\sigma _{\mu_{2}}^{(i_{2})}\,.\,\cdots\,.\,\frac{1}{2}\sigma _{\mu_{k}}^{(i_{k})} + \underset{i_{1}=i_{k}}{\mathlarger{\mathlarger{\sum}}_{i_{1}\neq i_{2}\neq\cdots\neq i_{k-1}=1}^{N}}\frac{1}{4}\sigma _{\mu_{1}}^{(i_{1})}\sigma _{\mu_{k}}^{(i_{1})}\,.\,\frac{1}{2}\sigma _{\mu_{2}}^{(i_{2})}\,.\,\cdots\,.\,\frac{1}{2}\sigma _{\mu_{k-1}}^{(i_{k-1})}\nonumber\\
    &+ \underset{i_{2}=i_{k}}{\mathlarger{\mathlarger{\sum}}_{i_{1}\neq i_{2}\neq\cdots\neq i_{k-1}=1}^{N}}\frac{1}{2}\sigma _{\mu_{1}}^{(i_{1})}\,.\,\frac{1}{4}\sigma _{\mu_{1}}^{(i_{2})}\sigma _{\mu_{k}}^{(i_{2})}\,.\,\cdots\,.\,\frac{1}{2}\sigma _{\mu_{k-1}}^{(i_{k-1})} +\cdots + \underset{i_{k-1}=i_{k}}{\mathlarger{\mathlarger{\sum}}_{i_{1}\neq i_{2}\neq\cdots\neq i_{k-1}=1}^{N}}\frac{1}{2}\sigma _{\mu_{1}}^{(i_{1})}\,.\,\frac{1}{2}\sigma _{\mu_{2}}^{(i_{2})}\,.\,\cdots\,.\,\frac{1}{4}\sigma _{\mu_{k-1}}^{(i_{k-1})}\sigma _{\mu_{k}}^{(i_{k-1})}.
\end{align}
The Pauli matrices satisfy
\begin{equation}\label{product_pauli}
    \sigma_{\mu}\sigma_{\nu}=\delta _{\mu\nu}\mathbb{1}+i\varepsilon _{\mu\nu\lambda}\sigma _{\lambda},
\end{equation}
where $\varepsilon _{\mu\nu\lambda}$ denotes the Levi-Civita symbol which is defined as follows: $\varepsilon _{xyz}=1$ for even permutations of $(x, y, z)$, $-1$ for odd permutations, and $0$ if any index is repeated. Applying Eq.~(\ref{product_pauli}) in Eq.~(\ref{deriv-k_body-p1}), yields
\begin{align}\label{deriv-k_body-p2}
    &\Bigg(\mathlarger{\mathlarger{\sum}}_{i_{1}\neq i_{2}\neq\cdots\neq i_{k-1}=1}^{N}\frac{1}{2}\sigma _{\mu_{1}}^{(i_{1})}\,.\,\frac{1}{2}\sigma _{\mu_{2}}^{(i_{2})}\,.\,\cdots\,.\,\frac{1}{2}\sigma _{\mu_{k-1}}^{(i_{k-1})}\Bigg) \Bigg(\sum_{i_{k}=1}^{N}\frac{1}{2}\sigma _{\mu_{k}}^{(i_{k})}\Bigg) =\nonumber\\
    &\mathlarger{\mathlarger{\sum}}_{i_{1}\neq i_{2}\neq\cdots\neq i_{k}=1}^{N}\frac{1}{2}\sigma _{\mu_{1}}^{(i_{1})}\,.\,\frac{1}{2}\sigma _{\mu_{2}}^{(i_{2})}\,.\,\cdots\,.\,\frac{1}{2}\sigma _{\mu_{k}}^{(i_{k})} + \underset{i_{1}=i_{k}}{\mathlarger{\mathlarger{\sum}}_{i_{1}\neq i_{2}\neq\cdots\neq i_{k-1}=1}^{N}}\frac{1}{4}\Big(\delta _{\mu_{1}\mu_{k}}\mathbb{1}+i\varepsilon _{\mu_{1}\mu_{k}\mu '}\sigma _{\mu'}^{(i_{1})}\Big)\,.\,\frac{1}{2}\sigma _{\mu_{2}}^{(i_{2})}\,.\,\cdots\,.\,\frac{1}{2}\sigma _{\mu_{k-1}}^{(i_{k-1})}\nonumber\\
    &+ \underset{i_{2}=i_{k}}{\mathlarger{\mathlarger{\sum}}_{i_{1}\neq i_{2}\neq\cdots\neq i_{k-1}=1}^{N}}\frac{1}{2}\sigma _{\mu_{1}}^{(i_{1})}\,.\,\frac{1}{4}\Big(\delta _{\mu_{2}\mu_{k}}\mathbb{1}+i\varepsilon _{\mu_{2}\mu_{k}\mu '}\sigma _{\mu'}^{(i_{2})}\Big)\,.\,\cdots\,.\,\frac{1}{2}\sigma _{\mu_{k-1}}^{(i_{k-1})} +\cdots\nonumber\\ 
    &+ \underset{i_{k-1}=i_{k}}{\mathlarger{\mathlarger{\sum}}_{i_{1}\neq i_{2}\neq\cdots\neq i_{k-1}=1}^{N}}\frac{1}{2}\sigma _{\mu_{1}}^{(i_{1})}\,.\,\cdots\,.\,\frac{1}{4}\Big(\delta _{\mu_{k-1}\mu_{k}}\mathbb{1}+i\varepsilon _{\mu_{k-1}\mu_{k}\mu '}\sigma _{\mu'}^{(i_{k-1})}\Big)\nonumber\\
    &=\mathlarger{\mathlarger{\sum}}_{i_{1}\neq i_{2}\neq\cdots\neq i_{k}=1}^{N}\frac{1}{2}\sigma _{\mu_{1}}^{(i_{1})}\,.\,\frac{1}{2}\sigma _{\mu_{2}}^{(i_{2})}\,.\,\cdots\,.\,\frac{1}{2}\sigma _{\mu_{k}}^{(i_{k})}\nonumber\\
    &+ \frac{\delta _{\mu_{1}\mu_{k}}}{4}\underset{i_{1}=i_{k}}{\mathlarger{\mathlarger{\sum}}_{i_{1}\neq i_{2}\neq\cdots\neq i_{k-1}=1}^{N}}\frac{1}{2}\sigma _{\mu_{2}}^{(i_{2})}\,.\,\cdots\,.\,\frac{1}{2}\sigma _{\mu_{k-1}}^{(i_{k-1})}+\frac{i}{2}\varepsilon _{\mu_{1}\mu_{k}\mu '}\underset{i_{1}=i_{k}}{\mathlarger{\mathlarger{\sum}}_{i_{1}\neq i_{2}\neq\cdots\neq i_{k-1}=1}^{N}}\frac{1}{2}\sigma _{\mu'}^{(i_{1})}\,.\,\frac{1}{2}\sigma _{\mu_{2}}^{(i_{2})}\,.\,\cdots\,.\,\frac{1}{2}\sigma _{\mu_{k-1}}^{(i_{k-1})}\nonumber\\       
    &+ \frac{\delta _{\mu_{2}\mu_{k}}}{4}\underset{i_{2}=i_{k}}{\mathlarger{\mathlarger{\sum}}_{i_{1}\neq i_{2}\neq\cdots\neq i_{k-1}=1}^{N}}\frac{1}{2}\sigma _{\mu_{1}}^{(i_{1})}\,.\,\cdots\,.\,\frac{1}{2}\sigma _{\mu_{k-1}}^{(i_{k-1})}+\frac{i}{2}\varepsilon _{\mu_{2}\mu_{k}\mu '}\underset{i_{2}=i_{k}}{\mathlarger{\mathlarger{\sum}}_{i_{1}\neq i_{2}\neq\cdots\neq i_{k-1}=1}^{N}}\frac{1}{2}\sigma _{\mu_{1}}^{(i_{1})}\,.\,\frac{1}{2}\sigma _{\mu '}^{(i_{2})}\,.\,\cdots\,.\,\frac{1}{2}\sigma _{\mu_{k-1}}^{(i_{k-1})}+\cdots\nonumber\\
    &+ \frac{\delta _{\mu_{k-1}\mu_{k}}}{4}\underset{i_{k-1}=i_{k}}{\mathlarger{\mathlarger{\sum}}_{i_{1}\neq i_{2}\neq\cdots\neq i_{k-1}=1}^{N}}\frac{1}{2}\sigma _{\mu_{1}}^{(i_{1})}\,.\,\cdots\,.\,\frac{1}{2}\sigma _{\mu_{k-2}}^{(i_{k-2})}+\frac{i}{2}\varepsilon _{\mu_{k-1}\mu_{k}\mu '}\underset{i_{k-1}=i_{k}}{\mathlarger{\mathlarger{\sum}}_{i_{1}\neq i_{2}\neq\cdots\neq i_{k-1}=1}^{N}}\frac{1}{2}\sigma _{\mu_{1}}^{(i_{1})}\,.\,\frac{1}{2}\sigma _{\mu_{2}}^{(i_{2})}\,.\,\cdots\,.\,\frac{1}{2}\sigma _{\mu'}^{(i_{k-1})}.
\end{align}
In Eq.~(\ref{deriv-k_body-p2}), one can perform a summation of $i_{j}$ $(j\in\{1, 2, \cdots, k\})$ in terms containing the factor $\delta_{\mu_{j}\mu_{k}}$. Moreover, given the definition of $k$-body interaction terms in Eq.~(\ref{k-body-cor}), each summation term in Eq.(\ref{deriv-k_body-p2}) can be written as lower-order interaction terms as follows
\begin{align}\label{deriv-k_body-p3}
    (k-1)!S_{\mu_{1}\mu_{2}\cdots\mu_{k-1}}S_{\mu_{k}} = k!S_{\mu_{1}\mu_{2}\cdots\mu_{k}} &+ \frac{(k-2)!(N-k+2)}{4} \sum_{i=1}^{k-1}\delta_{\mu_{i}\mu_{k}}S_{\mu_{1}\cdots\hat{\mu}_{i}\cdots\mu_{k-1}} \notag\\
    & + \frac{i(k-1)!}{2}\sum_{i=1}^{k-1}\varepsilon_{\mu_{i}\mu_{k}\mu'}S_{\mu_{1}\cdots\mu_{i}\rightarrow\mu'\cdots\mu_{k-1}},
\end{align}
where $\mu_{i}$ is omitted from the list in $S_{\mu_{1}\cdots\hat{\mu}_{i}\cdots\mu_{k-1}}$ and $S_{\mu_{1}\cdots\mu_{i}\rightarrow\mu'\cdots\mu_{k-1}}$ states that $\mu_{i}$ is replaced by $\mu'$. Hence, the $k$-body interaction terms can be given as a combination of ($k-1$)- and ($k-2$)-body interaction terms
\begin{equation}\label{k-body-cor-main}
    S_{\mu_{1}\mu_{2}\cdots\mu_{k}}=\frac{1}{k}S_{\mu_{1}\mu_{2}\cdots\mu_{k-1}} S_{\mu_{k}}- \frac{(N-k+2)}{4k(k-1)} \sum_{i=1}^{k-1} \delta_{\mu_{i}\mu_{k}} S_{\mu_{1}\cdots\hat{\mu}_{i}\cdots\mu_{k-1}} - \frac{i}{2k} \sum_{i=1}^{k-1} \varepsilon_{\mu_{i}\mu_{k}\mu'} S_{\mu_{1}\cdots\mu_{i}\rightarrow\mu'\cdots\mu_{k-1}}.
\end{equation}

To showcase the application of Eq.~(\ref{k-body-cor-main}), we consider
\begin{itemize}
    \item k=2
        \begin{align}
            S_{xx}&=\frac{1}{2}S_{x}^{2}-\frac{N}{8}\mathbb{1}\\
            S_{xz}&=\frac{1}{2}S_{x}S_{z}+\frac{i}{4}S_{y}
        \end{align}
    \item k=3
        \begin{align}
            S_{xxx}&=\frac{1}{3}S_{xx}S_{x}-\frac{N-1}{12}S_{x}\\
            S_{xxy}&=\frac{1}{3}S_{xx}S_{y}-\frac{i}{3}S_{xz}
        \end{align}                
    \item k=4
        \begin{align}
            S_{xxxx}&=\frac{1}{4}S_{xxx}S_{x}-\frac{N-2}{16}S_{xx}\\
            S_{xxxz}&=\frac{1}{4}S_{xxx}S_{z}+\frac{3i}{8}S_{xxy}
        \end{align}        
\end{itemize}

\subsection{Permutation symmetry}\label{sec:PI}
In order to efficiently compute the $k$-body interaction terms, one can apply the permutationally invariant property of Eq.~(\ref{k-body-cor-main}). This means that for each $k$, one solely needs to compute
\begin{equation}\label{k-body-cor-main-PI-form}
    S_{x^{a}y^{b}z^{c}},
\end{equation}
where non-negative integers $a,b,c$ satisfy $a+b+c=k$.
For notational simplicity, we will henceforth 
denote these by $S_{abc}$.
In other words, instead of computing $3^{k}$ terms, one needs to compute $\binom{k+2}{2}=(k+2)(k+1)/2$ terms for each $k$. For example, for 
\begin{itemize}
    \item $k=1: S_{100}, S_{010}, S_{001}$,
    \item $k=2: S_{200}, S_{020}, S_{002}, S_{110}, S_{101}, S_{011}$,
    \item $k=3: S_{300}, S_{030}, S_{003}, S_{210}, S_{201}, S_{021}, S_{120}, S_{102}, S_{012}, S_{111}$,
    \item $k=4: S_{400}, S_{040}, S_{004}, S_{310}, S_{301}, S_{220}, S_{202}, S_{211}, S_{121}, S_{112}, S_{130}, S_{103}, S_{022}, S_{031}, S_{013}$,
\end{itemize}

should be computed. By applying the permutation symmetry, Eq.~(\ref{k-body-cor-main}) can be recast as follows
\begin{align}\label{k-body-cor-main-PI-form-formula-3}
        S_{abc}&=\frac{1}{k}S_{ab(c-1)}S_{001}- \frac{(N-k+2)(c-1)}{4k(k-1)} S_{ab(c-2)}- \frac{i}{2k}(a \varepsilon_{xzy} S_{(a-1)(b+1)(c-1)}+b \varepsilon_{yzx} S_{(a+1)(b-1)(c-1)})\nonumber\\
        &=\frac{1}{k}S_{ab(c-1)}S_{001}- \frac{(N-k+2)(c-1)}{4k(k-1)} S_{ab(c-2)}+\frac{i}{2k}(aS_{(a-1)(b+1)(c-1)}-b S_{(a+1)(b-1)(c-1)}),
\end{align}
where $k=a+b+c$. Substituting $a=b=0$, yields 
\begin{equation}\label{Sz-k}
    S_{00c}=\frac{1}{c}S_{00(c-1)}S_{001}-\frac{(N-c+2)}{4c}S_{00(c-2)}.  
\end{equation}

For the case where $c=0$, Eq.~(\ref{k-body-cor-main}) is given by
\begin{align}\label{k-body-cor-main-PI-form-formula-2}
        S_{ab0}&=\frac{1}{k}S_{a(b-1)0}S_{010}- \frac{(N-k+2)(b-1)}{4k(k-1)} S_{a(b-2)0}- \frac{i}{2k}a \varepsilon_{xyz} S_{(a-1)(b-1)1}\nonumber\\
        &=\frac{1}{k}S_{a(b-1)0}S_{010}- \frac{(N-k+2)(b-1)}{4k(k-1)} S_{a(b-2)0}- \frac{ia}{2k}S_{(a-1)(b-1)1}.  
\end{align}
Finally,
\begin{align}\label{k-body-cor-main-PI-form-formula-1}
        S_{a00}&=\frac{1}{a}S_{(a-1)00}S_{100}- \frac{(N-a+2)}{4a} S_{(a-2)00}.
\end{align}

 For numerical precision, it is advisable to use the normalized version of the interaction terms, $\tilde{S}_{abc} =  {N\choose k}^{-1}2^kS_{abc}$ in order to avoid large matrix elements during the dynamical programming routine.  

\subsection{Computing interaction terms in the symmetric subspace}\label{sec: computing_symm}
We now compute the $k$-body interaction terms, Eq.~(\ref{k-body-cor-main}), in the symmetric subspace. The operators $S_{x}, S_{y}, S_{z}$ in Eq.~(\ref{commute-rel}) are the fundamental representation of $\text{SU(2)}$. 
In the notation introduced in the previous section, these one-body operators correspond to
$S_{x} = S_{100}$, $S_{y} = S_{010}$, and $S_{z} = S_{001}$.
The higher irreducible representation of $\text{SU(2)}$ can be constructed from the product representation of spin-$\frac{1}{2}$. For example, in the case of four spin-$\frac{1}{2}$ systems, the tensor product space is given by
\begin{align}\label{irr_rep}
    \mathbb{C}^{2}\otimes\mathbb{C}^{2}\otimes\mathbb{C}^{2}\otimes\mathbb{C}^{2}&=(\mathbb{C}^{1}\oplus\mathbb{C}^{3})\otimes\mathbb{C}^{2}\otimes\mathbb{C}^{2}\nonumber\\
    &=(\mathbb{C}^{2}\oplus\mathbb{C}^{2}\oplus\mathbb{C}^{4})\otimes\mathbb{C}^{2}\nonumber\\
    &=\mathbb{C}^{1}\oplus\mathbb{C}^{2}\oplus\mathbb{C}^{1}\oplus\mathbb{C}^{2}\oplus\mathbb{C}^{2}\oplus\mathbb{C}^{5},
\end{align}
where $\mathbb{C}^{1}$, $\mathbb{C}^{2}$, $\mathbb{C}^{3}$, $\mathbb{C}^{4}$, and $\mathbb{C}^{5}$ correspond to spin-$0$, spin-$\frac{1}{2}$, spin-$1$, spin-$\frac{3}{2}$, and spin-$2$ representation, respectively. The spin-$S$ representation corresponds to ($2S+1$)-dimensional vector space in which $m=\{-S, -S+1, \cdots, S-1, S\}$. In this representation, the basis states are the common eigenstates of $S_{z}$ and $S^{2}$ which is defined as
\begin{equation}\label{Ssquared}
    S^{2}=S_{x}^{2}+S_{y}^{2}+S_{z}^{2}.
\end{equation}
Hence
\begin{align}
    S_{z}\ket{S,m} &= m\ket{S,m},\label{Szeigen}\\
    S^{2}\ket{S,m} &= S(S+1)\ket{S,m}\label{S2eigen}.
\end{align}
Introducing \textit{raising} and \textit{lowering} operator $S_{+}$ and $S_{-}$, respectively, as follows
\begin{equation}\label{ladder_op}
    S_{\pm}\ket{S,m}=\sqrt{(S\mp m)(S\pm m +1)}\ket{S,m\pm 1},
\end{equation}
leads to 
\begin{align}
    S_{x}&=\frac{1}{2}(S_{+}+S_{-})\label{def_Sx_ladder},\\
    S_{y}&=\frac{1}{2i}(S_{+}-S_{-})\label{def_Sy_ladder}.
\end{align}
This representation simplifies the computation of $k$-body interaction terms in the symmetric subspace. The symmetric subspace can be spanned by the Dicke basis $\{\ket{D_{N}^{0}}, \ket{D_{N}^{1}}, \cdots, \ket{D_{N}^{N}}\}$ in $(N+1)$ dimensional subspace. This subspace corresponds to the irreducible representation of spin-$S$ with dimension $(2S+1)$. Therefore, any arbitrary quantum state, $\ket{\psi_0}$, in the symmetric subspace can be written as follows
\begin{equation}\label{state-sym-sub-spin}
    \ket{\psi}=\sum _{n=0}^{N}C_{n}\ket{D_{N}^{n}}\equiv\sum_{m=-N/2}^{N/2}\alpha _{m}\ket{\frac{N}{2}, m},
\end{equation}
where $C_{n}, \alpha _{m}\in\mathbb{C}$. For instance in the case of $N=2$:
\begin{align}
   \frac{1}{2}\otimes\frac{1}{2}&\overset{(\ref{irr_rep})}{=} 0\oplus1\nonumber\\
   &~=
\begin{pmatrix}
  0 & \vdots & &\\
  \cdots & \cdots &\cdots&\\
   & \vdots & \Large 1_{3\times3}&\\
   &\vdots & &
\end{pmatrix},
\end{align}
for $S=1$, $m=\{-1,0,+1\}$. The symmetric subspace can be constructed with the following bases
\begin{align*}
    \ket{D_{2}^{0}}&=\ket{00},\nonumber\\
    \ket{D_{2}^{1}}&=\frac{1}{\sqrt{2}}(\ket{10}+\ket{01}),\nonumber\\
    \ket{D_{2}^{2}}&=\ket{11},
\end{align*}
which is a $3$-dimensional space. This shows that the spin-$1$ representation can be utilized to represent the symmetric subspace.

The next step is finding the representation of one-body interaction terms, Eq.~(\ref{one-body-cor}), in spin-$S$ representation (or symmetric subspace). These are given by
\begin{align}
    \bra{\frac{N}{2},m'}S_{x}\ket{\frac{N}{2},m} &=\frac{1}{2}\Big[\sqrt{\Big(\frac{N}{2}-m\Big)\Big(\frac{N}{2}+m +1\Big)}\delta_{m',m+1}+\sqrt{\Big(\frac{N}{2}+m\Big)\Big(\frac{N}{2}-m+1\Big)}\delta_{m',m-1}\Big],\label{Sx_sym_rep}\\
    \bra{\frac{N}{2},m'}S_{y}\ket{\frac{N}{2},m}  &=\frac{1}{2i}\Big[\sqrt{\Big(\frac{N}{2}-m\Big)\Big(\frac{N}{2}+m +1\Big)}\delta_{m',m+1}-\sqrt{\Big(\frac{N}{2}+m\Big)\Big(\frac{N}{2}-m +1\Big)}\delta_{m',m-1}\Big],\label{Sy_sym_rep}\\
    \bra{\frac{N}{2},m'}S_{z}\ket{\frac{N}{2},m}  &=m~\delta_{m',m},\label{Sz_sym_rep}
\end{align}
where $m', m\in \{-N/2, -N/2 +1,\cdots, N/2 -1, N/2 \}$. For $N=2$, the spin-$1$ representations of one-body interaction terms are derived as follows
\begin{align}
    S_{x}&=\frac{1}{\sqrt{2}}\left(\begin{array}{ccc}
        0&1&0\\ 
        1&0&1\\
        0&1&0\\
    \end{array}\right)\label{Sx_sym_rep_1},\\
    S_{y}&=\frac{1}{i\sqrt{2}}\left(\begin{array}{ccc}
        0&1&0\\ 
        -1&0&1\\
        0&-1&0\\
    \end{array}\right)\label{Sy_sym_rep_1},\\
    S_{z}&=\left(\begin{array}{ccc}
        1&0&0\\ 
        0&0&0\\
        0&0&-1\\
    \end{array}\right)\label{Sz_sym_rep_1}.
\end{align}
Substituting Eqs. (\ref{Sx_sym_rep_1}, \ref{Sy_sym_rep_1},\ref{Sz_sym_rep_1}) in Eq.~(\ref{Ssquared}), yields
\begin{equation}\label{Ssquared_sym_rep_1}
    S^{2}=\left(\begin{array}{ccc}
        2&0&0\\ 
        0&2&0\\
        0&0&2\\
    \end{array}\right),
\end{equation}
which is compatible with Eq.~(\ref{S2eigen}) for $S=1$
\begin{equation}\label{Ssquared_sym_rep}
    \bra{\frac{N}{2},m'}S^{2}\ket{\frac{N}{2},m}=S(S+1)\delta_{m',m}.
\end{equation}
To estimate the computational complexity of constructing the operators $S_{abc}$, we first note that the total number of distinct permutation-invariant operators at fixed $k$ is equal to $\binom{k+2}{2}$, and sum over all correlators up to $k$, yields $\sum _{k}\binom{k+2}{2}=O(k^{3})$, since the indices satisfy $a+b+c=k$. Furthermore, all calculations are performed within the symmetric subspace, whose dimension scales as $N+1$. Consequently, the matrix representation of each operator has dimension $(N+1)\times(N+1)$, and dense matrix operations scale as $O((N+1)^3)=O(N^3)$. Combining these observations, the overall complexity for recursively constructing all correlators scales as
\begin{equation}\label{complexity_recursive_S}
    O(k^{3}N^{3}).
\end{equation}

\section{Computing the computational complexity of the QFI in the symmetric subspace}\label{Com_Complex_QFI_symm_sub}
In order to determine the computational complexity of evaluating the QFI in Eq.~(\ref{QFI-symmetric-subspace}), we analyze separately the cost of computing $\mathpzc{C}_{mn}$ and $\alpha _{m}$. For $\mathpzc{C}_{mn}$, we note the following:
\begin{itemize}
    \item The summation $\sum_{\vec{\omega}~\vdash N}$ runs over all compositions of $N$ into four non-negative integers $(\bar{\omega}_{0},\bar{\omega}_{1},\bar{\omega}_{2},\bar{\omega}_{3})$ satisfying $\bar{\omega}_{0}+\bar{\omega}_{1}+\bar{\omega}_{2}+\bar{\omega}_{3}=N$. The total number of such compositions is $\binom{N+3}{3}=O(N^{3})$.
    \item The two Kronecker delta functions, \(\delta(\bar{\omega}_{2}+\bar{\omega}_{3}-m)\) and \(\delta(\bar{\omega}_{1}+\bar{\omega}_{3}-n)\), impose the constraints \(\bar{\omega}_{2}=m-\bar{\omega}_{3}\) and \(\bar{\omega}_{1}=n-\bar{\omega}_{3}\). Using the normalization condition \(\bar{\omega}_{0}=N-\bar{\omega}_{1}-\bar{\omega}_{2}-\bar{\omega}_{3}\), the summation reduces to a single free variable \(\bar{\omega}_{3}\).
    \item The allowed range of \(\bar{\omega}_{3}\) is determined by non-negativity of all components, yielding \(\bar{\omega}_{3} \in [\max(0,m+n-N),\, \min(m,n)]\). Therefore, the number of terms in the sum scales as \(O(\min(m,n)) \leqslant O(N)\).
    \item The multinomial coefficient
    \[\frac{N!}{\bar{\omega}_{0}!\bar{\omega}_{1}!\bar{\omega}_{2}!\bar{\omega}_{3}!},\]
    the powers \(\mathcal{U}_{ij}^{\bar{\omega}_{k}}(\theta)\), and the prefactor \(\frac{1}{\sqrt{\binom{N}{m}\binom{N}{n}}}\) can be evaluated in constant time, \(O(1)\), assuming standard precomputation or reuse of intermediate quantities.
\end{itemize}
Since the cost per term is $O(1)$ and the number of terms scales as $O(N)$, the total computational complexity of evaluating a single coefficient $\mathpzc{C}_{mn}$ is $O(N)$.
To determine the computational complexity of $\alpha_m$, we note that $\alpha_m$ denotes the expansion coefficient of the ground state in the symmetric subspace. These coefficients are obtained from the exact diagonalization of the Hamiltonian constructed from the $k$-body correlators. The total computational complexity can therefore be separated into two parts:
\begin{itemize}
    \item the complexity of recursively constructing the operators $S_{abc}$, which from Eq.~(\ref{complexity_recursive_S}) scales as $O(k^{3}N^{3})$,
    \item the complexity of computing the ground state within the symmetric subspace. Since the Hamiltonian acts on a Hilbert space of dimension $N+1$, exact diagonalization scales as $O(N^{3})$.
\end{itemize}
Since the QFI in Eq.~(\ref{QFI-symmetric-subspace}) involves a triple summation, each index running over $(N+1)$ values, the summation itself costs $O(N^{3})$ iterations. At each iteration:
\begin{itemize}
    \item the coefficients $\mathpzc{C}_{mn}$ and their derivatives are evaluated on the fly at a cost of $O(N)$ each,
    \item the coefficients $\alpha_{m}$ are precomputed via exact diagonalization at a one-time cost of $O(N^{3})$ and stored, so accessing $\alpha_{m}^{*}\alpha_{n}$ at each iteration costs $O(1)$.
\end{itemize}
Therefore, evaluating the triple summation in Eq.~(\ref{QFI-symmetric-subspace}) costs $O(N^{3})\times O(N)=O(N^{4})$. Combining all contributions, the total computational complexity breaks down as follows:
\begin{enumerate}
    \item constructing the $S_{abc}$ operators recursively costs $O(k^{3}N^{3})$,
    \item obtaining the initial-state coefficients $\alpha_{m}$ via exact diagonalization costs $O(N^{3})$,
    \item evaluating the triple summation costs $O(N^{4})$.
\end{enumerate}
Therefore, the total computational complexity of evaluating the QFI in Eq.~(\ref{QFI-symmetric-subspace}) is
\begin{equation}\label{computational_complexity}
    O(k^{3}N^{3})+O(N^{3})+O(N^{4})=O(N^{4}+k^{3}N^{3}).
\end{equation}


\section{QFI and the energy gap}

Here we present additional plots of the energy gap versus the QFI for different interaction orders $k$ (see Fig.~\ref{GapVSQFI} of the main text).

\begin{figure}[h]
  \centering
  \subfloat[ $k=4$]{\includegraphics[width=0.5\linewidth]{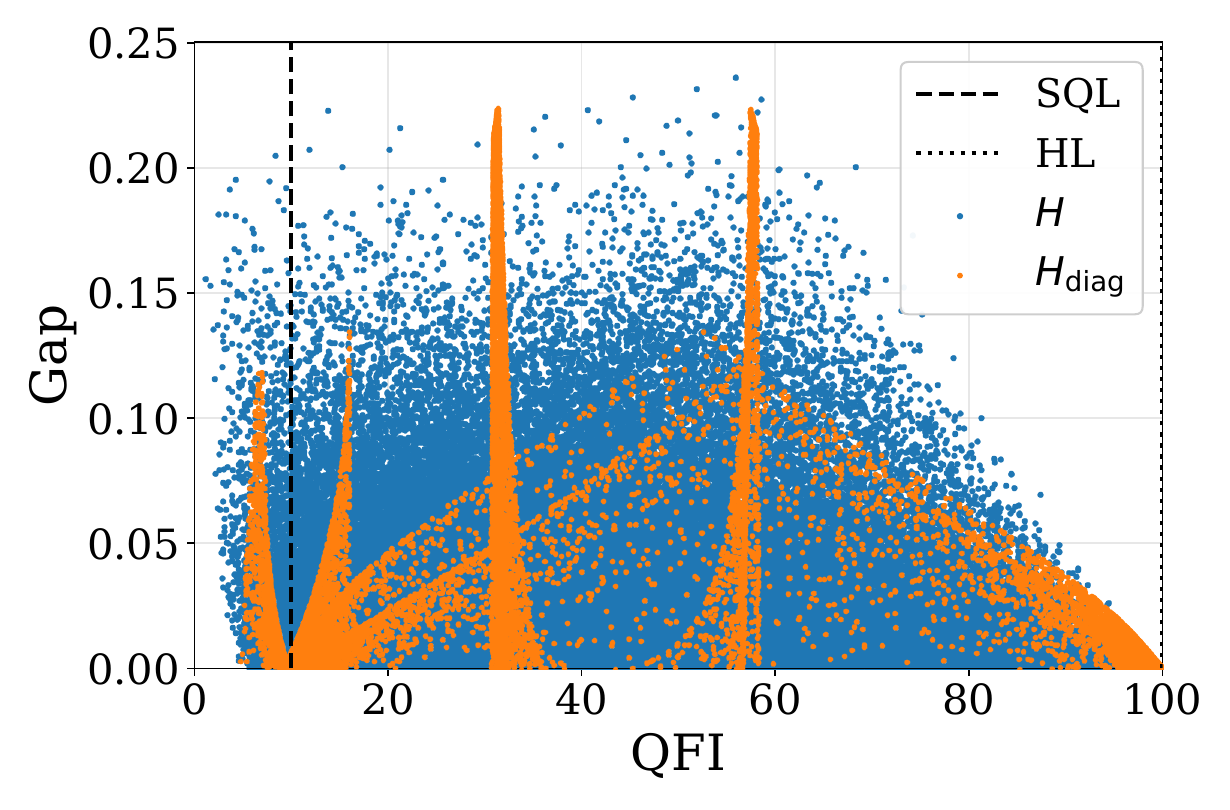}}
  \subfloat[ $k=6$]{\includegraphics[width=0.5\linewidth]{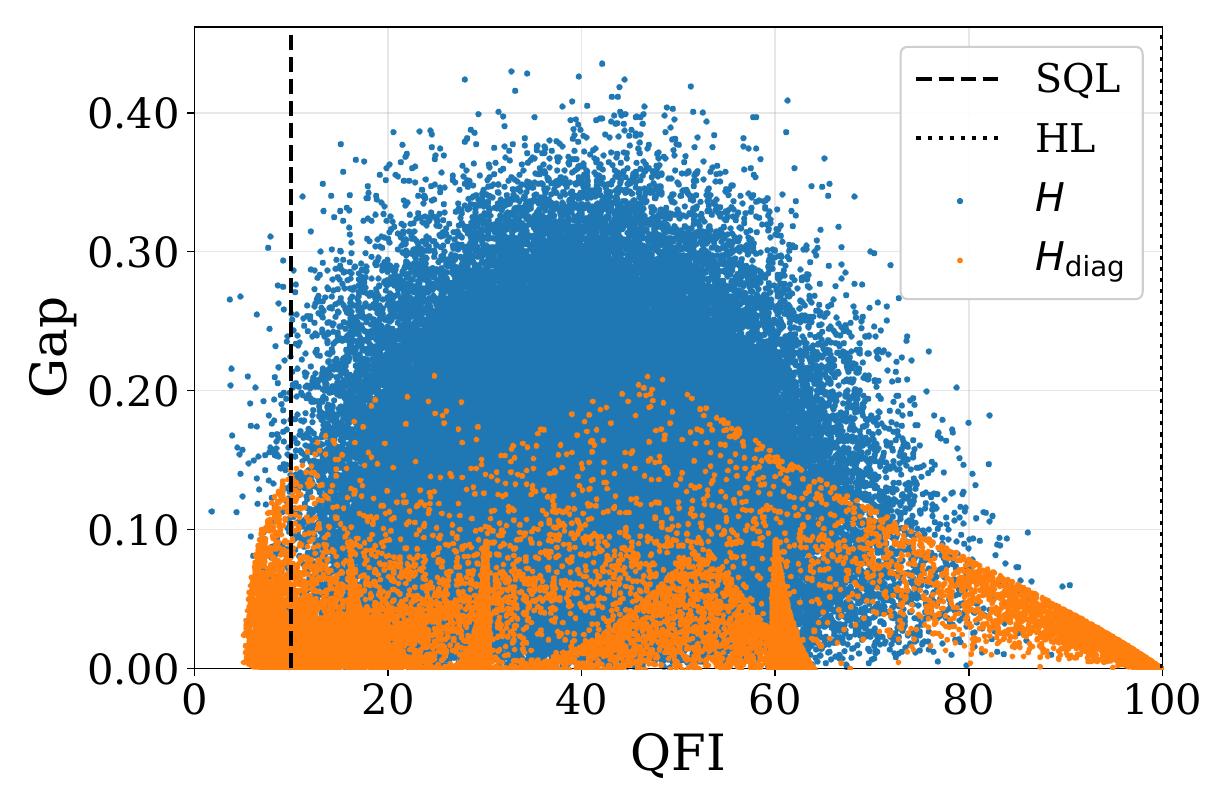}}
  
  \subfloat[ $k=8$]{\includegraphics[width=0.5\linewidth]{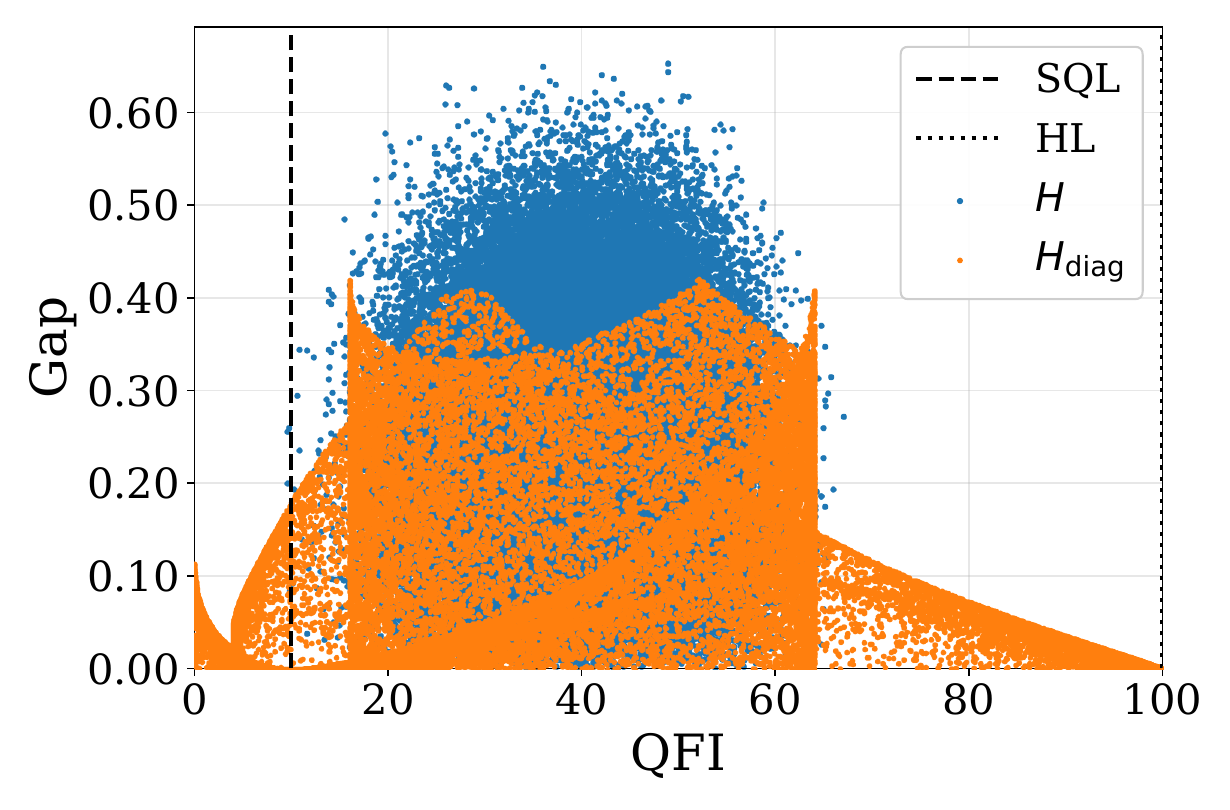}}
   \subfloat[ $k=10$]{\includegraphics[width=0.5\linewidth]{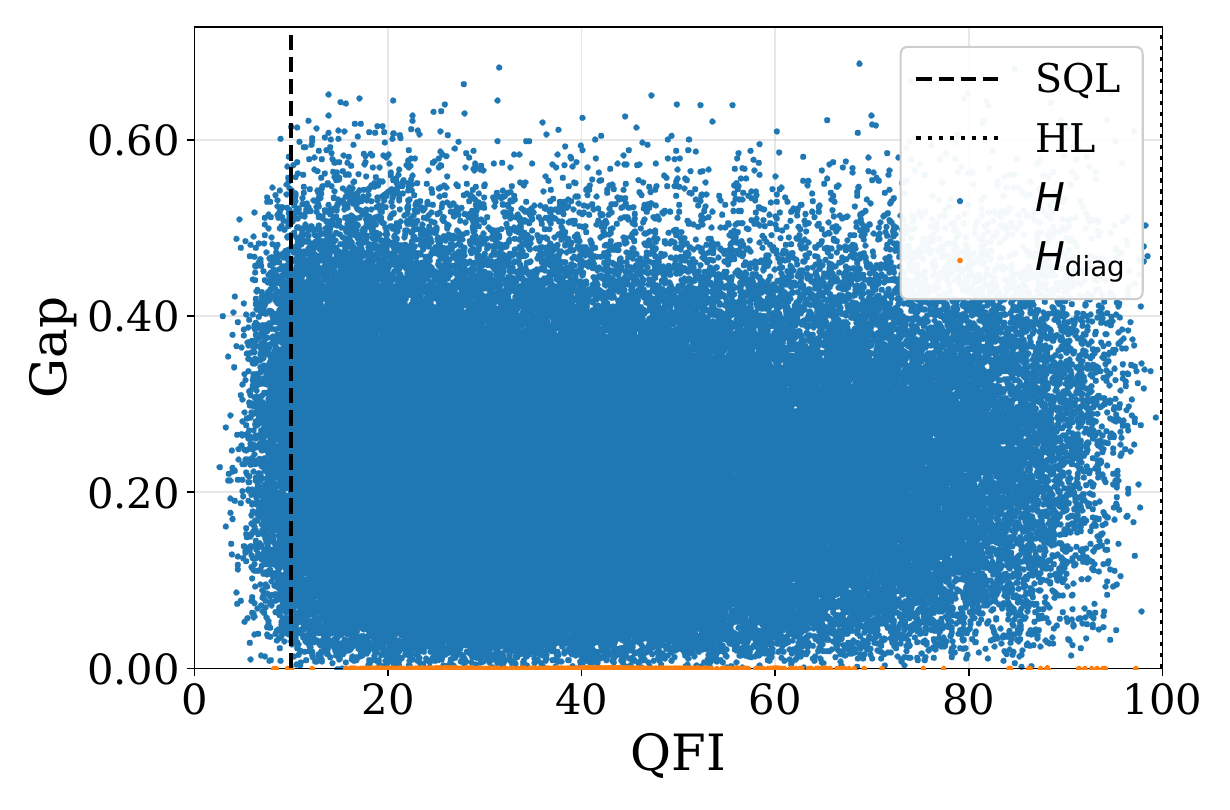}}
   \caption{Energy gap and corresponding QFI for $10^5$ ground states of random Hamiltonians involving interactions between $k$ spins for $N=10$ particles (blue dots). Orange dots correspond to Hamiltonians with only nonzero coefficients, $\Gamma_{k00},\Gamma_{0k0},  \Gamma_{00k}$.}
   \label{fig:QFIGapEndMatter}
\end{figure}

We notice that diagonal Hamiltonians (orange dots) cease to be optimal for $k>2$. On the other hand, increasing the order of the interactions leads to a larger gap (notice $y$-axis scale), which facilitates the preparation of states with high QFI (see rightmost part of the plots). In the limiting case $k=N=10$, diagonal Hamiltonians are degenerated, and, as expected, the tradeoff between the QFI and the gap imposed by the order of the interactions is lifted.


\section{Parameter-dependent generators}
To demonstrate that our approach applies to parameter-dependent generators, we consider $G_\theta = \cos(\theta)\sigma _{z}+\sin(\theta)\sigma _{x}$. Using Eq.~\eqref{QFI_PDH_Variance_upper_bound}, we derive the $\theta$-independent bound $\qfi{[\ket{\psi_\theta}]}\leqslant [2\sin(1)N]^2$, and we demonstrate that this bound is saturated by our method,  see Fig.~\ref{PDH_E1-optimized} below.  

\begin{figure}[h!]
    \centering
    \subfloat[ Average and fluctuations of the QFI ]{\includegraphics[width=0.6\linewidth]{ 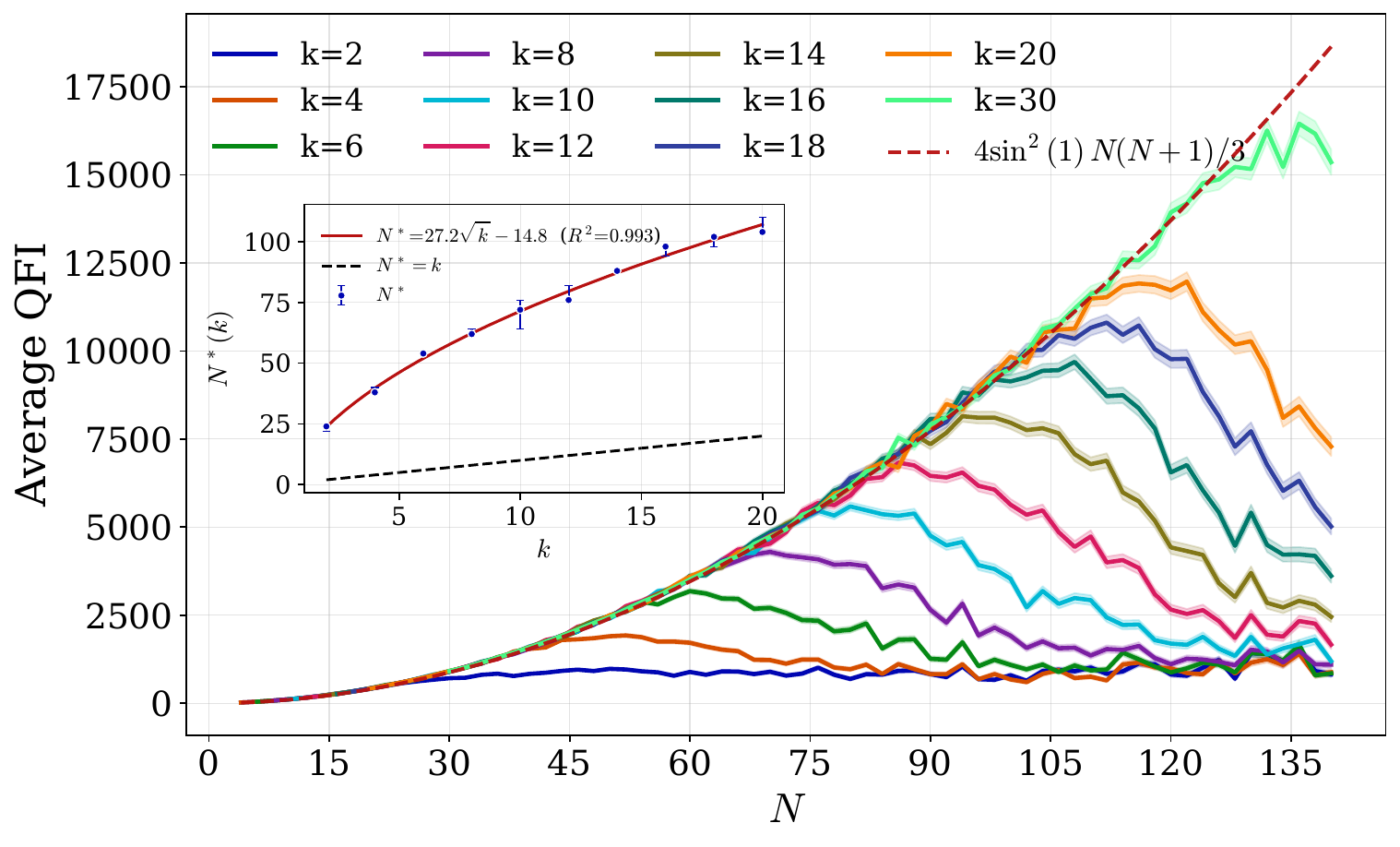}} \\
    \subfloat[ Distribution of the QFI when $k=4$]{\includegraphics[width=0.6\linewidth]{ 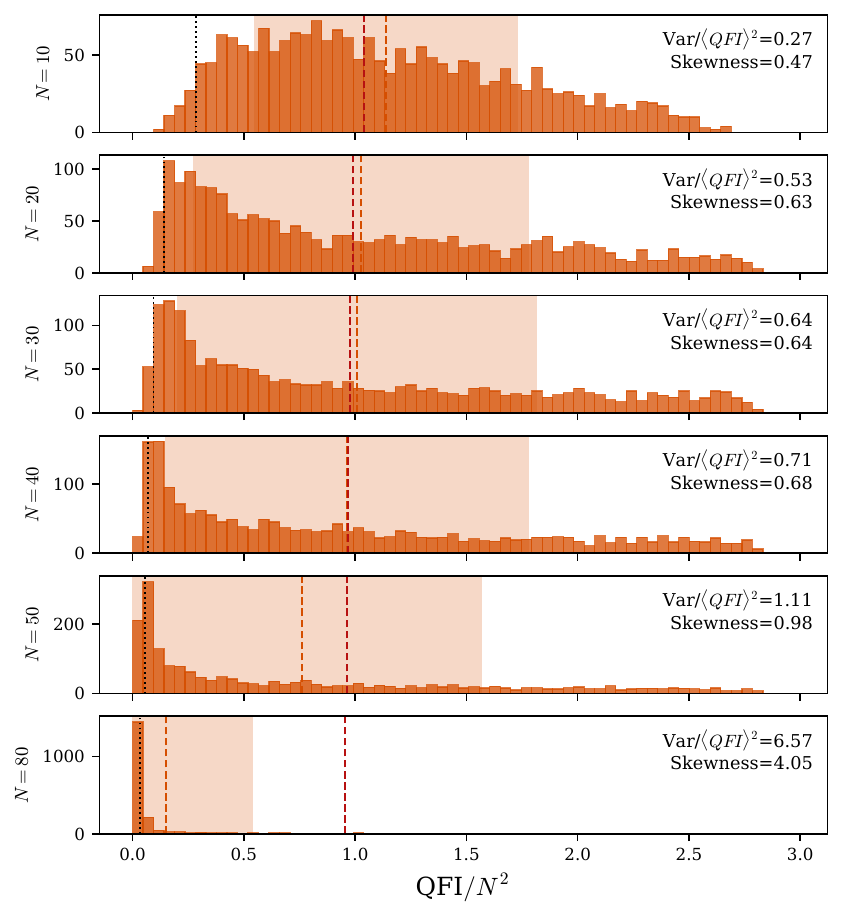}}
    \caption{(a) The QFI as a function of $N$ for the encoding generator $G_\theta = \cos(\theta)\sigma _{z}+\sin(\theta)\sigma _{x}$ (cf. Fig.~\ref{H_X-random_sampling}) for random $\theta$. The probe states are the ground states of $H$, constructed with randomly sampled coefficients $\Gamma _{abc}$ drawn as in the main text, for various $k$-body interactions; solid curves are sample averages over 2000 realizations, shaded regions $\pm$SEM, and the dashed line the Haar value $4\sin^{2}(1)N(N+1)/3$. Inset: $N^{*}(k)$ and the fit $N^{*}=27.2\sqrt{k}-14.8$ ($R^{2}=0.993$). (b) Histograms of $\mathrm{QFI}/N^{2}$ for $k=4$ and $N=10,20,30,40,50,80$, with the same conventions as Fig.~\ref{H_X-random_sampling}. (b) The orange dashed
    line marks the sample average $\langle \mathrm{QFI}\rangle$, the shaded
    band spans one standard deviation about the mean,
    $\langle \mathrm{QFI}\rangle \pm \sigma$.
    The dark-red dashed line marks the Haar value and the dotted line the SQL.}
    \label{PDH_E1-random_sampling}
\end{figure}

Since the QFI is no longer continuous in this case~\cite{rezakhani_continuity_2019}, the results of Ref.~\cite{oszmaniec_random_2016} are not directly applicable. Instead, we employ our method to characterize how random ground states perform as probes relative to the upper bound. Figure.~\ref{PDH_E1-random_sampling} shows the outcome of randomly sampling the coefficients in Eq.~(\ref{H_sys}) of the main text. These results reveal that few-body interactions already suffice to reach the Heisenberg-like scaling for moderate system sizes.
\begin{figure}[h!]
    \centering
    \includegraphics[width=0.8\textwidth]{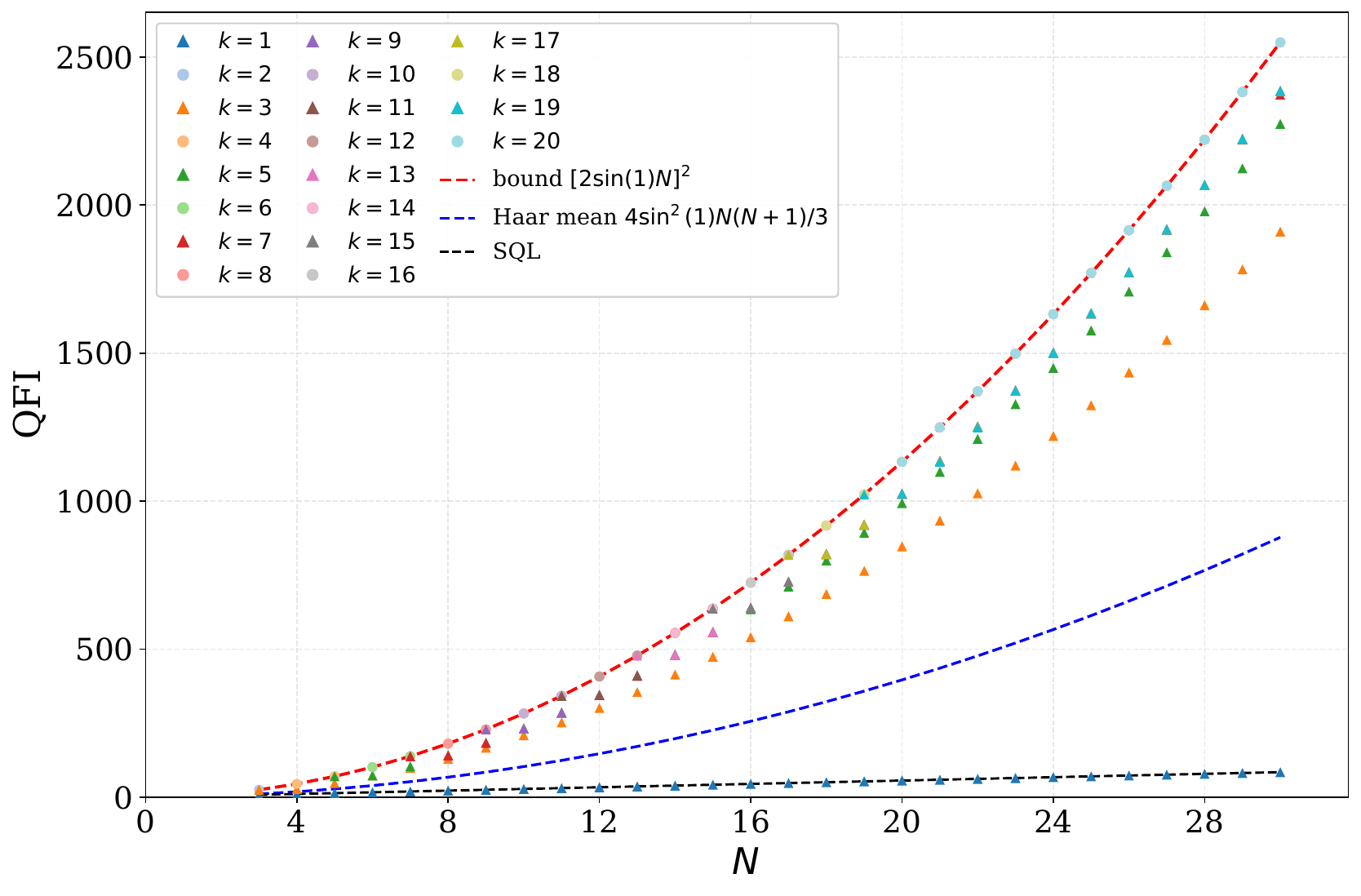}
    \caption{The QFI as a function of $N$ for the case where the encoding Hamiltonian is $G_\theta = \cos(\theta)\sigma _{z}+\sin(\theta)\sigma _{x}$. By optimization, for every even $k$ and for $k=N$ the bound $[2\sin(1)N]^{2}$ of Eq.~\eqref{bound} (red dashed) is saturated.
    }
    \label{PDH_E1-optimized}
\end{figure}

To finish the section, let us comment on the saturation of the bound Eq.~\eqref{bound}. Direct computation shows that states of the form $\ket{\psi_\theta} = (\ket{\boldsymbol{\lambda}_M(\theta)} +\ket{\boldsymbol{\lambda}_m(\theta)})/\sqrt{2}$, where $\ket{\boldsymbol{\lambda}_{M/m}(\theta)}$ are respectively the max/min eigenvectors of $\mathbf{K}_{\theta}$ saturate the bound. Moreover, since $\mathbf{K}_{\theta} = \sum_{i=1}^N K_{\theta}^{(i)}$, where $K_{\theta}^{(i)}$ only acts nontrivially on qubit $i$,  $\ket{\boldsymbol{\lambda}_{M/m}(\theta)} = \ket{\lambda_{M/m}(\theta)}^{\otimes N}$, where $\ket{{\lambda}_{M/m}(\theta)}$ are respectively the max/min eigenvectors of $K_{\theta}$.

\begin{figure}[h!]
    \centering
    \includegraphics[width=0.90\textwidth]{ 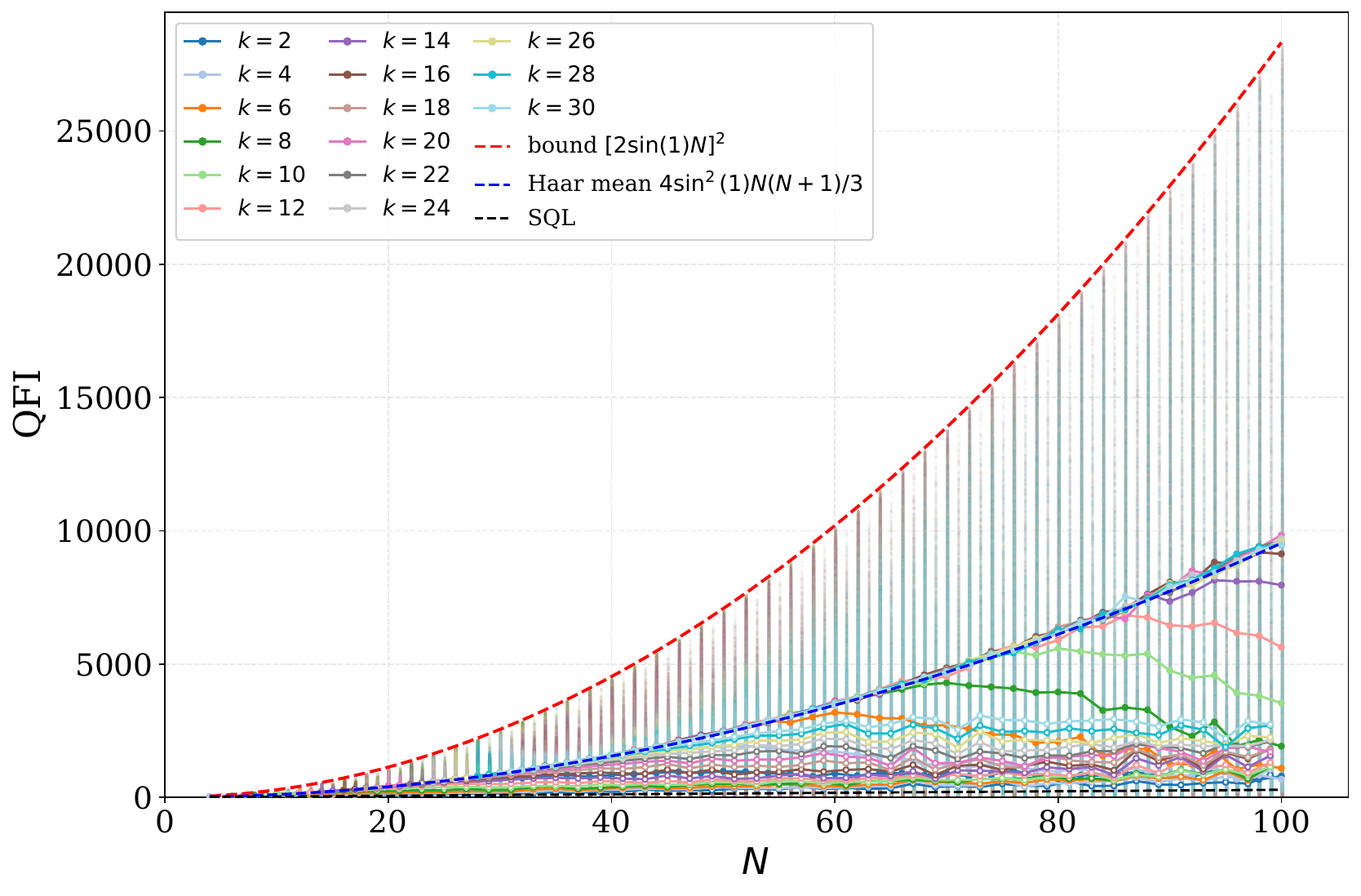}
    \caption{
    The QFI of the ground states of random Hamiltonians as a function of $N$ (even and odd, $N\leqslant100$) for the encoding generator $G_\theta = \cos(\theta)\sigma _{z}+\sin(\theta)\sigma _{x}$, for even interaction orders $k\leqslant30$ (dots: 2000 random instances per $(N,k)$; lines: sample averages, filled markers for even $N$ and open markers for odd $N$). Dashed lines (red, blue, and dark): the bound $[2\sin(1)N]^{2}$, the Haar value $4\sin^{2}(1)N(N+1)/3$, and the SQL, respectively. 
    }
    \label{PDH_E1-random_sampling_sat}
\end{figure}

Figure~\ref{PDH_E1-random_sampling_sat} presents the results of utilizing our method for the parameter-dependent case $G_\theta = \cos(\theta)\sigma _{z}+\sin(\theta)\sigma _{x}$.
At first glance, Fig.~\ref{PDH_E1-random_sampling_sat} may suggest that the upper bound is saturated only for even values of $N$. However, this apparent behavior arises due to the randomness of the sampled coefficients and does not reflect a fundamental limitation. As we mentioned, in Fig.~\ref{PDH_E1-optimized}, we go beyond random sampling by directly optimizing the QFI and find that for all values of $N$, there exist corresponding system's Hamiltonian whose ground states saturate the upper bound.

\newpage
\clearpage

\section{Minimal sets of interactions}
Here you can find the minimal set of $k$-body interactions to produce random Hamiltonians with Haar distributed eigenstates.

\renewcommand{\arraystretch}{1.2}
\setlength{\tabcolsep}{6pt}
\begin{table}[!htbp]
	\caption{{Sufficient set of interactions to span the space of Hamiltonians}}
	\label{tab:bases}
	\begin{tabular}{|c|c|c|p{9cm}|}
	 \hline
\textbf{$N$} & \textbf{Final Rank $(N+1)^2$} &
\textbf{Subset Size} & \textbf{Sufficient set of interaction terms} \\
\hline
3 & 16 & 16 &
$\mathbb{1}, S_{001}, S_{010}, S_{100}, S_{002}, S_{011}, S_{020}, S_{101},  S_{110}, S_{003}, S_{012}, S_{021},$ \\
& & & 
$ S_{030}, S_{102}, S_{111}, S_{120}$
\\ \hline

4 & 25 & 25 &
$\mathbb{1}, S_{001}, S_{010}, S_{100}, S_{002}, S_{011}, S_{020}, S_{101}, S_{110}, S_{003}, S_{012}, S_{021}, $\\
& & & 
$ S_{030}, S_{102}, S_{111}, S_{120}, S_{004}, S_{013}, S_{022}, S_{031}, S_{040}, S_{103}, S_{112}, S_{121}, $\\
& & & 
$S_{130}$
\\ \hline

5 & 36 & 36 &
$\mathbb{1}, S_{001}, S_{010}, S_{100}, S_{002}, S_{011}, S_{020}, S_{101}, S_{110}, S_{003}, S_{012}, S_{021}, $\\
& & &
$S_{030}, S_{102}, S_{111}, S_{120}, S_{004}, S_{013}, S_{022}, S_{031}, S_{040}, S_{103}, S_{112}, S_{121}, $\\
& & &
$ S_{130}, S_{005}, S_{014}, S_{023}, S_{032}, S_{041}, S_{050}, S_{104}, S_{113}, S_{122},
S_{131}, S_{140}$ 
\\ \hline

6 & 49 & 49 &
$\mathbb{1}, S_{001}, S_{010}, S_{100}, S_{002}, S_{011}, S_{020}, S_{101}, S_{110}, S_{003}, S_{012}, S_{021}, $\\
& & &
$S_{030}, S_{102}, S_{111}, S_{120},
S_{004}, S_{013}, S_{022}, S_{031}, S_{040}, S_{103}, S_{112}, S_{121},$\\
& & &
$S_{130}, S_{005}, S_{014}, S_{023}, S_{032}, S_{041}, S_{050}, S_{104}, S_{113}, S_{122}, S_{131}, S_{140}, $\\
& & &
$S_{006}, S_{015}, S_{024}, S_{033}, S_{042}, S_{051}, S_{060}, S_{105}, S_{114}, S_{123}, S_{132}, S_{141}, $\\
& & &
$ S_{150} $
\\ \hline

10 & 121 & 121 &
$\mathbb{1}, S_{0 0 1}, S_{0 1 0}, S_{1 0 0}, S_{0 0 2}, S_{0 1 1}, S_{0 2 0}, S_{1 0 1}, S_{1 1 0},
S_{0 0 3}, S_{0 1 2}, S_{0 2 1}, $ \\
& & &
$ S_{0 3 0}, S_{1 0 2}, S_{1 1 1}, S_{1 2 0},
S_{0 0 4}, S_{0 1 3}, S_{0 2 2}, S_{0 3 1}, S_{0 4 0}, S_{1 0 3}, S_{1 1 2}, S_{1 2 1}, $\\
& & &
$ S_{1 3 0},
S_{0 0 5}, S_{0 1 4}, S_{0 2 3}, S_{0 3 2}, S_{0 4 1}, S_{0 5 0}, S_{1 0 4}, S_{1 1 3}, S_{1 2 2},
S_{1 3 1}, S_{1 4 0}, $\\
& & &
$ S_{0 0 6}, S_{0 1 5}, S_{0 2 4}, S_{0 3 3}, S_{0 4 2}, S_{0 5 1}, S_{0 6 0}, S_{1 0 5}, S_{1 1 4},
S_{1 2 3}, S_{1 3 2}, S_{1 4 1}, $\\
& & &
$ S_{1 5 0},
S_{0 0 7}, S_{0 1 6}, S_{0 2 5}, S_{0 3 4}, S_{0 4 3}, S_{0 5 2}, S_{0 6 1}, S_{0 7 0}, S_{1 0 6}, S_{1 1 5}, S_{1 2 4},  $\\
& & &
$ S_{1 3 3}, S_{1 4 2}, S_{1 5 1}, S_{1 6 0}, S_{0 0 8}, S_{0 1 7}, S_{0 2 6}, S_{0 3 5}, S_{0 4 4}, S_{0 5 3}, S_{0 6 2}, S_{0 7 1}, $\\
& & &
$ S_{0 8 0}, S_{1 0 7}, S_{1 1 6}, S_{1 2 5}, S_{1 3 4}, S_{1 4 3}, S_{1 5 2}, S_{1 6 1}, S_{1 7 0}, S_{0 0 9}, S_{0 1 8}, S_{0 2 7}, $\\
& & &
$S_{0 3 6}, S_{0 4 5}, S_{0 5 4}, S_{0 6 3}, S_{0 7 2}, S_{0 8 1}, S_{0 9 0}, S_{1 0 8}, S_{1 1 7}, S_{1 2 6}, S_{1 3 5}, S_{1 4 4}, $\\
& & &
$ S_{1 5 3}, S_{1 6 2}, S_{1 7 1}, S_{1 8 0},
S_{0 \,0\, 10}, S_{0 1 9}, S_{0 2 8}, S_{0 3 7}, S_{0 4 6}, S_{0 5 5}, S_{0 6 4}, S_{0 7 3},$\\
& & &
$  S_{0 8 2}, S_{0 9 1}, S_{0 \,10\, 0}, S_{1 0 9}, S_{1 1 8}, S_{1 2 7}, S_{1 3 6}, S_{1 4 5}, S_{1 5 4}, S_{1 6 3}, S_{1 7 2}, S_{1 8 1},$\\
& & &
$ S_{1 9 0}$
\\ \hline
\end{tabular}
\end{table}

\end{document}